\newif\ifconfver
\confvertrue        
\ifconfver
    \documentclass[10pt,twocolumn,twoside]{IEEEtran}
\else
    \documentclass[11pt,draftcls,onecolumn]{IEEEtran}
\fi

\usepackage{cite}
\usepackage{graphicx}
\usepackage{psfrag}
\usepackage{subfigure}
\usepackage{url}
\usepackage{stfloats}
\usepackage{amsfonts,amssymb,amsmath,amsbsy,bm,paralist,theorem,cite,ifthen,color}
\usepackage{array}
\usepackage{calc}
\usepackage{subfig}
\usepackage{longtable}
\usepackage{multirow}
\usepackage{algorithmic,algorithm}
\usepackage{graphics,booktabs,epsfig,enumerate}

\newtheorem{Lemma}{Lemma}

\newtheorem{Theorem}{Theorem}
\newtheorem{Definition}{Definition}

\newtheorem{Remark}{Remark}
\theorembodyfont{\rmfamily}

\begin{document}

\def\blue{\color{blue}}
\def\red{\color{red}}
\definecolor{orange}{RGB}{255,107,0}
\def\orange{\color{orange}}
\definecolor{green}{RGB}{0,180,80}
\def\green{\color{green}}

\newcommand\Ab{\mathbf{A}}
\newcommand\Fb{\mathbf{F}}
\newcommand\Pb{\mathbf{P}}
\newcommand\Qb{\mathbf{Q}}
\newcommand\Rb{\mathbf{R}}
\newcommand\Ub{\mathbf{U}}
\newcommand\Vb{\mathbf{V}}
\newcommand\Wb{\mathbf{W}}

\newcommand\ab{\bm{a}}
\newcommand\eb{\bm{e}}
\newcommand\hb{\bm{h}}
\newcommand\vb{\bm{v}}
\newcommand\wb{\bm{w}}
\newcommand\xb{\bm{x}}

\newcommand\Lambdab{\bm{\Lambda}}

\newcommand\lambdab{\bm{\lambda}}
\newcommand\mub{\bm{\mu}}
\newcommand\nub{\bm{\nu}}

\newcommand\Cplx{\mathbb{C}}
\newcommand\zerob{\mathbf{0}}

\newcommand\tr{\mathrm{Tr}}
\newcommand\rank{\mathrm{rank}}

\markboth{Submitted to IEEE Transactions on Signal Processing, May
2014}{Submitted to IEEE Transactions on Signal Processing, May 2014}

\title{Multicell Coordinated
Beamforming with Rate Outage Constraint--Part I: Complexity
Analysis} \ifconfver \else {\linespread{1.1} \rm \fi

\author{\vspace{0.5cm}Wei-Chiang Li$^\ast$, Tsung-Hui Chang, and Chong-Yung
Chi
\thanks{Part of this work will be presented in IEEE ICASSP 2014
\cite{Li2014}. The work is supported by the National Science
Council, R.O.C., under Grants NSC-102-2221-E-007-019-MY3 and NSC
102-2221-E-011-005-MY3.}
\thanks{Wei-Chiang Li and Chong-Yung Chi are with Institute of
Communications Engineering \& Department of Electrical Engineering,
National Tsing Hua University, Hsinchu, Taiwan 30013, R.O.C. E-mail:
weichiangli@gmail.com, cychi@ee.nthu.edu.tw}
\thanks{Tsung-Hui Chang is with Department of Electronic and Computer
Engineering, National Taiwan University of Science and Technology,
Taipei, Taiwan 10607, R.O.C. E-mail: tsunghui.chang@ieee.org.} }


\maketitle

\vspace{-\baselineskip}
\begin{abstract}
This paper studies the coordinated beamforming (CoBF) design in the
multiple-input single-output interference channel, assuming only
channel distribution information given a priori at the transmitters.
The CoBF design is formulated as an optimization problem that
maximizes a predefined system utility, e.g., the weighted sum rate
or the weighted max-min-fairness (MMF) rate, subject to constraints
on the individual probability of rate outage and power budget. While
the problem is non-convex and appears difficult to handle due to the
intricate outage probability constraints, so far it is still unknown
if this outage constrained problem is computationally tractable. To
answer this, we conduct computational complexity analysis of the
outage constrained CoBF problem. Specifically, we show that the
outage constrained CoBF problem with the weighted sum rate utility
is intrinsically difficult, i.e., NP-hard. Moreover, the outage
constrained CoBF problem with the weighted MMF rate utility is also
NP-hard except the case when all the transmitters are equipped with
single antenna. The presented analysis results confirm that
efficient approximation methods are indispensable to the outage
constrained CoBF problem.
\ifconfver \else
\\\\
\noindent {\bfseries EDICS}: SAM-BEAM, SPC-CCMC, SPC-INTF, SPC-APPL
\fi
\end{abstract}

%

\ifconfver \else \IEEEpeerreviewmaketitle} \fi

\ifconfver \else \newpage \fi

\section{Introduction}\label{sec:intr}

Coordinated transmission has been recognized as a promising approach
to improving the system performance of wireless cellular networks
\cite{Lee_2012_ComMag}. According to the level of cooperation, the
coordinated transmission can be roughly classified into two
categories, i.e., \textit{MIMO cooperation} and \textit{interference
coordination} \cite{Gesbert10JSAC}. In MIMO cooperation, the
transmitters, e.g., base stations (BSs), cooperate in data
transmission by sharing all the channel state information (CSI) and
data signals. In interference coordination, the BSs only coordinate
in the transmission strategies for mitigating the inter-cell
interference. Compared with MIMO cooperation, interference
coordination requires the CSI to be shared only, and hence induces
less overhead on the backhaul network \cite{Bjornson11TSP}. A common
model for studying interference coordination is the interference
channel (IFC) \cite{Annapureddy2011}, where multiple transmitters
simultaneously communicate with their respective receivers over a
common frequency band, and thus interfere with each other.

In this paper, we consider a multiple-input single-output (MISO)
IFC, wherein the transmitters are equipped with multiple antennas
while the receivers are equipped with single antenna. Moreover, we
are interested in the \textit{coordinated beamforming} (CoBF) design
problem; that is, the transmitters coordinate with each other to
optimize their transmit beamforming vectors. A typical formulation
of the CoBF design problem is to maximize a system utility function,
e.g., the sum rate, proportional fairness rate, harmonic mean rate
or the max-min-fairness (MMF) rate, assuming that the transmitters
have perfect CSI. It turns out that the CoBF problems are in general
difficult optimization problems. Specifically, it has been shown in
\cite{Liu_11} that, except for the MMF rate \cite{Liu_13}, the CoBF
problem for the sum rate, proportional fairness rate and harmonic
mean rate are NP-hard in general, implying that they cannot be
efficiently solved in general. Due to this fact, a significant
amount of research efforts has been devoted to the development of
reliable and efficient methods for handling the CoBF problems. For
example, the works
\cite{Jorswieck08,Mochaourab_11,Zakhour_09,Bjornson_2012_Pareto}
characterize the optimal beamforming structure in order to reduce
the dimension of exhaustive search. Global optimization algorithms
were also developed in
\cite{MAPEL,Jorswieck2010_POA,Utschick2012_POA,Zhang_2012_POA} but
are only efficient when the number of users is small. Another branch
of works focus on suboptimal but computationally efficient
approximation algorithms; see
\cite{Liu_11,Schmidt_09,Zakhour_09,Zhang_Cui_2010,Kim_2011,Shi2011_IteMMSE,Nguyen_11,Hong2012,Weeraddana_2013}.

The works mentioned above all have assumed that the transmitters
have perfect CSI. However, in practical wireless scenarios,
especially in mobile channels, it is difficult for the transmitters
to acquire accurate CSI due to time-varying channels. In contrast,
the statistical distribution of the channel, i.e., the channel
distribution information (CDI), can remain static in a relatively
long period of time and thus is easier to obtain compared to the
CSI. However, with only CDI at the transmitters, the transmission
would suffer from rate outage; that is, the instantaneous channel
may not reliably support the data transmission. In view of this,
outage-aware CoBF designs, which constrain the probability of rate
outage to be low, have attracted extensive attention recently; see,
e.g., \cite{Lindblom_11,Park2012,Li_13} for the outage constrained
utility maximization problem, \cite{Kandukuri02,Ghosh_10} for the
outage constrained power minimization problem and
\cite{Kandukuri02,Tan_2011,Huang_2012_GLOBECOM} for the outage
balancing problem. One essential fact that is commonly observed in
these works is that the outage constrained CoBF problems are usually
nonconvex and appear much more difficult to handle than their
perfect CSI counterparts \cite{Liu_11,Liu_13}, due to the
complicated probabilistic outage constraints. Therefore, most of the
works \cite{Park2012,Li_13,Huang_2012_GLOBECOM} have concentrated on
developing efficient approximation algorithms. However, unlike the
perfect CSI case where the complexity status of various CoBF
problems has been identified \cite{Liu_11,Liu_13}, it is still not
clear if the outage constrained CoBF problems are computational
tractable or, instead, are intrinsically difficult, i.e., NP-hard
\cite{Karp2010_21NPComplete}.

Our interest in this paper lies in characterizing the computational
complexity status of the outage constrained CoBF problem.
Specifically, we consider the outage constrained (weighted) sum rate
maximization (SRM) CoBF problem and the outage constrained
(weighted) MMF CoBF problem, which respectively maximize the
(weighted) sum rate and (weighted) MMF rate under rate outage
constraints and individual power constraints. We analytically show
that the outage constrained SRM problem is NP-hard in general,  and
that the outage constrained MMF problem is polynomial-time solvable
when each of the transmitters is equipped with only one antenna but
is NP-hard when each of the transmitters is equipped with at least
two antennas. The NP-hardness of the outage constrained SRM problem
is established by a polynomial-time reduction from the NP-hard
Max-Cut problem \cite{Karp2010_21NPComplete}, i.e., each problem
instance of the Max-Cut problem can be transformed to a problem
instance of the outage constrained SRM problem in polynomial time;
while the NP-hardness of the outage constrained MMF problem in the
MISO scenario is established by a polynomial-time reduction from the
3-satisfiability (3-SAT) problem, which is known to be NP-complete
\cite{Karp2010_21NPComplete}. The proposed analysis about the
NP-hardness of MMF problem can also be analogously applied to prove
that the outage balancing problem studied in \cite{Tan_2011} is
NP-hard when each of the transmitters has at least two antennas. The
complexity analysis results further motivate the development of
efficient approximation algorithms for handling the outage
constrained CoBF problem; see \cite{Li_14_alg}.

{\bf Synopsis:} In Section \ref{sec:sys_mod_prob_state}, we present
the system model and problem formulations. The complexity analyses
for the outage constrained SRM problem and outage constrained MMF
problem are presented in Section
\ref{sec:WSRM_complexity_analysis_proof} and Section
\ref{sec:WMRM_complexity_analysis_proof}, respectively. Finally, we
draw the conclusions in Section \ref{sec:conclusion}.

{\bf Notation:} The set of $n$-dimensional real vectors, complex
vectors and complex Hermitian matrices are denoted by
$\mathbb{R}^n$, $\mathbb{C}^n$ and $\mathbb{H}^n$, respectively. The
set of non-negative real vectors and the set of
positive-semidefinite Hermitian matrices are denoted by
$\mathbb{R}_+^n$ and $\mathbb{H}_+^n$, respectively. The
superscripts `$T$' and `$H$' represent the matrix transpose and
conjugate transpose, respectively. We denote $\|\cdot\|$ and
$\lceil\cdot\rceil$ as the vector Euclidean norm and ceiling
function, respectively. $\mathbf{A}\succeq\zerob$ means that matrix
$\mathbf{A}$ is positive semidefinite. We use the expression
$\mathbf{x}\sim\mathcal{CN}(\bm{\mu},\mathbf{Q})$ if $\mathbf{x}$ is
circularly symmetric complex Gaussian distributed with mean
$\bm{\mu}$ and covariance matrix $\mathbf{Q}$. We denote
$\exp(\cdot)$ (or simply $e^{(\cdot)}$) as the exponential function,
while $\ln(\cdot)$ and $\Pr\{\cdot\}$ represent the natural
logarithm function and the probability function, respectively. For a
variable $a_{ik}$, where the domain of subscript pair $i,k$ are
clear from context, $\{a_{ik}\}$ denotes the set of all $a_{ik}$
with the subscript pair $i,k$ covering all the possible values over
its domain, and $\{a_{ik}\}_k$ denotes the set of all $a_{ik}$ with
the first subscript equal to $i$. The sets $\{a_{ik}\}_{k{\ne}j}$
and $\{a_{ik}\}_{(i,k)\ne(j,\ell)}$ are defined by the set
$\{a_{ik}\}_k$ excluding $a_{ij}$ and the set $\{a_{ik}\}$ excluding
$a_{j\ell}$, respectively.

\section{System Model and Problem Statement}\label{sec:sys_mod_prob_state}

We consider a MISO IFC with $K$ pairs of transmitters and receivers
(i.e., $K$ users). Each transmitter is equipped with $N_t$ antennas
and each receiver is equipped with single antenna. Transmit
beamforming is assumed for the communication between each
transmitter and its intended receiver. Let
$s_i\sim\mathcal{CN}(0,1)$ denote the message for the $i$th user,
and $\wb_i\in\Cplx^{N_t}$ denote the corresponding beamforming
vector. We assume a frequency flat channel model, and the channel
vector between transmitter $i$ and receiver $k$ is modeled as
$\hb_{ik}\sim\mathcal{CN}(\zerob,\Qb_{ik})$, where
$\Qb_{ik}\succeq\zerob$ is the channel covariance matrix. The
received signal for receiver $i$ is then given by
\begin{equation}\label{received signal}
x_i =\hb_{ii}^H\wb_is_i+\sum_{k=1,k\neq{i}}^K\hb_{ki}^H\wb_ks_k+n_i,
\end{equation}
where $n_i\sim\mathcal{CN}(0,\sigma_i^2)$ is the additive noise at
receiver $i$ with variance $\sigma_i^2>0$. Under the assumption that
single-user detection is used by each receiver, the instantaneous
achievable rate (in bits/sec/Hz) of the $i$th user can be expressed
as
\begin{equation}\label{eq:achievable_rate_CSI}
r_i\!\left(\{\hb_{ki}\}_k,\{\wb_k\}\right)=\log_2\!\left(1+\frac{\left|\hb_{ii}^H\wb_i\right|^2}{\sum_{k\neq{i}}\left|\hb_{ki}^H\wb_k\right|^2\!+\!\sigma^2_i}\right).
\end{equation}

To enhance the overall system performance, a typical formulation of
the CoBF design is to optimize the beamforming vectors
$\{\wb_k\}_{k=1}^K$ of all the $K$ users so as to maximize a
performance measuring system utility function, e.g., the information
sum rate, under power constraints \cite{Liu_11}, which can be
mathematically expressed as
\begin{subequations}\label{UMX_CSI}
\begin{align}
\max_{\substack{\wb_i\in\Cplx^{N_t},R_i\ge0,\\i=1,\dots,K}}~&U(R_1,\dots,R_K)\label{UMX_CSI_a}\\
\text{subject to (s.t.)}~&R_i\le{r}_i(\{\hb_{ki}\}_k,\{\wb_k\}),~i=1,\dots,K,\label{UMX_CSI_b}\\
&\|\wb_i\|^2\le{P}_i,~i=1,\dots,K.\label{UMX_CSI_c}
\end{align}
\end{subequations}
Here, $R_1,\dots,R_K$ are the respective transmission rates of the
$K$ users, and $P_1,\dots,P_K$ are the associated power budgets. The
function $U(R_1,\dots,R_K)$ denotes the system utility. In this
paper, we are interested in two system utilities in particular,
i.e., the weighted sum rate, where $U(R_1,\dots,R_K)=\sum_{i=1}^K
\alpha_i R_i$ and the weighted min rate, where
$U(R_1,\dots,R_K)=\min_{i=1,\ldots,K} R_i/\alpha_i$. The
corresponding utility optimization problems are known as the SRM
problem and the MMF problem, respectively. These two problem
formulations represent two extremes of the tradeoff between system
throughput and the user fairness. For the SRM problem, one aims to
maximize the system throughput, but the transmission may be
dominated by a few of users. For the MMF problem, one places the
highest emphasis on user fairness, but the achieved system
throughput may not be as high. Interestingly, the complexity status
of solving these two problems are also very different. Specifically,
solving the SRM problem is NP-hard in general \cite{Luo_Zhang2008}
which means that the problem is unlikely to be solved in a
polynomial-time complexity; by contrast, the MMF problem is
polynomial-time solvable \cite{Liu_11}. Efficient approximation
algorithms for handling the SRM problem have been extensively
studied (see \cite{Liu_11,Zhang_2012_POA} and references therein).

However, to solve the CoBF design problem \eqref{UMX_CSI}, it is
required that perfect instantaneous CSI is available at the
transmitters, leading to enormous communication overhead. It is
hence more appropriate to assume that only statistical channel
information, i.e., the set of channel covariance matrices
$\{\Qb_{ik}\}$, is available at the transmitters. Under such
circumstances, reliable transmission cannot be guaranteed, and the
users may suffer from rate outage. Specifically, given any
transmission rate $R_i>0$, the outage event
$r_i(\{\hb_{ki}\}_k,\{\wb_k\})<R_i$ occurs with a non-zero
probability. It is therefore desirable to constrain the probability
of rate outage below a preassigned threshold. Let
$\epsilon_i\in(0,1)$ be the maximum tolerable rate-outage
probability for user $i$. To the end, we consider the following
outage constrained CoBF design problem \cite{Li_13}
\begin{subequations}\label{UMX_CDI_Pr}
\begin{align}
\max_{\substack{\wb_i\in\Cplx^{N_t},R_i\ge0,\\i=1,\dots,K}}~&U(R_1,\dots,R_K)\label{UMX_CDI_Pr_a}\\
\text{s.t.}~&\Pr\left\{r_i(\{\hb_{ki}\}_k,\{\wb_k\})<R_i\right\}\le\epsilon_i,\label{UMX_CDI_Pr_b}\\
&\|\wb_i\|^2{\le}P_i,~i=1,\dots,K.\label{UMX_CDI_Pr_c}
\end{align}
\end{subequations}
According to \cite{Li2011,Li_13}, the outage constraint
\eqref{UMX_CDI_Pr_b} can be explicitly expressed as
\begin{equation}\label{eq:outage_constraint_CLSFORM}
\rho_i\exp\!\left(\!\frac{(2^{R_i}\!-\!1)\sigma_i^2}{\wb_i^H\Qb_{ii}\wb_i}\!\right)\!\prod_{k{\ne}i}\!\left(\!1\!+\!\frac{(2^{R_i}\!-\!1)\wb_k^H\Qb_{ki}\wb_k}{\wb_i^H\Qb_{ii}\wb_i}\!\right)\!\le\!1,
\end{equation}
where $\rho_i\triangleq1-\epsilon_i$ for $i=1,\dots,K$ are the
satisfaction probabilities required in the downlink transmission.

Due to the complicated constraint
\eqref{eq:outage_constraint_CLSFORM}, solving the outage constrained
problem \eqref{UMX_CDI_Pr} seems more difficult than solving its
perfect CSI counterpart, i.e., problem \eqref{UMX_CSI}. However,
this intuitive observation is not mathematically precise. It is
hence of interest to investigate the complexity status of problem
\eqref{UMX_CDI_Pr}. In the ensuing sections, we study the complexity
of solving problem \eqref{UMX_CDI_Pr} with weighted sum rate and
minimum rate utilities, which correspond to the SRM and MMF
formulations, respectively. Our complexity analysis will demonstrate
that problem \eqref{UMX_CDI_Pr} is indeed more challenging.
Specifically, problem \eqref{UMX_CDI_Pr} is NP-hard not only for the
SRM formulation but also for the MMF formulation, while problem
\eqref{UMX_CSI} is at least polynomial-time solvable for the MMF
formulation \cite{Liu_11}. The obtained results about the complexity
of problem \eqref{UMX_CDI_Pr}, together with the corresponding
results in the literature about problem \eqref{UMX_CSI}, are
summarized in Table \ref{table:ex1margin} on the top of the next
page.

\begin{table*}[t]\centering\extrarowheight=2pt\vspace{-.2cm}
\caption{Summary of complexity analysis
results}\vspace{-0.2cm}\label{table:ex1margin}
\begin{center}
    \begin{tabular}{c|c|c|c|c}
    \toprule
    \multicolumn{1}{c}{}& \multicolumn{1}{c}{{\sf SRM} with CSI} & \multicolumn{1}{c}{{\sf SRM} with CDI} & \multicolumn{1}{c}{{\sf MMF} with CSI} & \multicolumn{1}{c}{{\sf MMF} with CDI}\\
    \multicolumn{1}{c}{}& \multicolumn{1}{c}{\cite[Theorem 1]{Luo_Zhang2008}} & \multicolumn{1}{c}{Theorem 1} & \multicolumn{1}{c}{\cite[Theorem 3.3]{Liu_11}} & \multicolumn{1}{c}{Theorems 2 \& 3}\\
    \bottomrule
    \multirow{2}{*}{$N_t=1$} &  NP-hard  & NP-hard  & \multirow{2}{*}{Polynomial-time solvable} & \multirow{2}{*}{Polynomial-time solvable} \\
    & (Max Indep. Set Problem) &  (Max-Cut Problem) & & \\
    \hline
    \multirow{2}{*}{$N_t>1$} & \multirow{2}{*}{NP-hard} & \multirow{2}{*}{NP-hard} & \multirow{2}{*}{Polynomial-time solvable} & NP-hard \\
    & & & & (3-SAT Problem)\\
    \toprule
    \end{tabular}
\end{center}
\hrulefill\vspace{-0.4cm}
\end{table*}

\section{Complexity Analysis for Outage constrained SRM problem}\label{sec:WSRM_complexity_analysis_proof}

In this section, we analyze the complexity status of the outage
constrained SRM problem, which can be written as\vspace{-.3cm}
\begin{subequations}\label{WSRM}
\begin{align}
\max_{\substack{\wb_i\in\Cplx^{N_t},R_i\ge0,\\i=1,\dots,K}}~&\sum_{i=1}^K \alpha_iR_i\label{WSRM_a}\\
\text{s.t.}~&\rho_i\exp\left(\frac{(2^{R_i}-1)\sigma_i^2}{\wb_i^H\Qb_{ii}\wb_i}\right)\notag\\
&~\times\!\prod_{k{\ne}i}\!\left(1\!+\!\frac{(2^{R_i}-1)\wb_k^H\Qb_{ki}\wb_k}{\wb_i^H\Qb_{ii}\wb_i}\right)\le1,\label{WSRM_b}\\
&\|\wb_i\|^2{\le}P_i,~i=1,\dots,K.\label{WSRM_c}
\end{align}
\end{subequations}
Specifically, we demonstrate that problem \eqref{WSRM} is NP-hard in
general. The following theorem makes our statement precise.

\begin{Theorem}\label{thm:WSRM_complexity}
The outage constrained SRM problem \eqref{WSRM} is NP-hard in the
number of users $K$, for all $N_t\ge1$.
\end{Theorem}

Theorem \ref{thm:WSRM_complexity} indicates that problem
\eqref{WSRM} is computationally intractable, like its perfect CSI
counterpart \eqref{UMX_CSI} with the weighted sum rate utility
\cite{Luo_Zhang2008}. While both of these two problems are NP-hard,
one should note that the techniques used for proving Theorem
\ref{thm:WSRM_complexity} are quite different from those used in
\cite{Luo_Zhang2008}. In \cite{Luo_Zhang2008}, it was shown that
problem \eqref{UMX_CSI} with the weighted sum rate utility is at
least as difficult as the maximum independent set problem (which is
known NP-complete) \cite{Karp2010_21NPComplete}. However, the same
idea is not applicable to the complexity analysis for problem
\eqref{WSRM}, due to the much more involved constraints
\eqref{WSRM_b}. Instead, we show in the next subsection that problem
\eqref{WSRM} is at least as difficult as the Max-Cut problem.

\subsection{Proof of Theorem \ref{thm:WSRM_complexity}}

Here, we show that problem \eqref{WSRM} is NP-hard even when
$N_t=1$, which implies that problem \eqref{WSRM} is NP-hard for the
general case of $N_t\ge1$. For $N_t=1$, the CoBF design problem
\eqref{WSRM} reduces to a coordinated power control problem.
Specifically, the MISO channel $\hb_{ki}\in\mathbb{C}^{N_t}$ reduces
to the single-input single-output (SISO) channel
$h_{ki}\in\mathbb{C}$, the channel covariance matrix
$\Qb_{ki}\in\mathbb{H}_+^{N_t}$ reduces to $Q_{ki}\in\mathbb{R}_+$,
and the beamformer $\wb_i$ reduces to the square root of the
transmit power $p_i$; thus, problem \eqref{WSRM} reduces
to\vspace{-.2cm}

\begin{subequations}\label{WSRM_pwr}
\begin{align}
\max_{\substack{p_i\in\mathbb{R},R_i\ge0,\\i=1,\dots,K}}~&\sum_{i=1}^K\alpha_iR_i\label{WSRM_Pwr_a}\\
\text{s.t.}~&\rho_i\exp\left(\frac{(2^{R_i}-1)\sigma_i^2}{Q_{ii}p_i}\right)\notag\\
&~\times\prod_{k{\ne}i}\left(1+\frac{(2^{R_i}-1)Q_{ki}p_k}{Q_{ii}p_i}\right)\le1,\label{WSRM_pwr_b}\\
&0 ~{\le}~ p_i~ {\le}~ P_i,~i=1,\dots,K.\label{WSRM_pwr_c}
\end{align}
\end{subequations}
We will show that the weighted Max-Cut problem, which is known to be
NP-hard \cite{Karp2010_21NPComplete}, is polynomial-time reducible
to problem \eqref{WSRM_pwr}, i.e., the Max-Cut problem is a special
instance of \eqref{WSRM_pwr}. For ease of the ensuing presentation,
the definition of the weighted Max-Cut problem is repeated as
follows.

\begin{Definition}\label{def:MAXCUT}
Consider an undirected and connected graph
$G=(\mathcal{V},\mathcal{E})$, where
$\mathcal{V}=\{1,\dots,|\mathcal{V}|\}$ denotes the set of vertices
in $G$, and
$\mathcal{E}=\{(i,j)|~i\in\mathcal{V},~j\in\mathcal{V},~i<j,~\text{and}~i,j~\text{are
connected in $G$}\}$ denotes the set of edges in $G$. Each edge
$(i,j)\in \mathcal{E}$ is assigned with a weight $w_{ij}>0$. A cut,
$\mathcal{E}(\mathcal{S}) \subseteq \mathcal{E}$, consists of the
set of edges crossing a subset $\mathcal{S}\subseteq\mathcal{V}$ and
its complement $\bar{\mathcal{S}}=\mathcal{V}\backslash\mathcal{S}$.
The weighted Max-Cut problem is formulated as
\begin{equation}\label{MAXCUT}
\max_{\mathcal{S}\subseteq\mathcal{V}}~\sum_{(i,j)\in\mathcal{E}(\mathcal{S})}w_{ij}.
\end{equation}
\end{Definition}

To build the connection between the Max-Cut problem and problem
\eqref{WSRM_pwr}, we consider an alternative formulation of
\eqref{WSRM_pwr}:

\begin{Lemma}\label{lemma:equivalent pwc}
Problem \eqref{WSRM_pwr} can be equivalently expressed as
    \begin{subequations}\label{WSRM_pwr_ctrl}
        \begin{align}
        \max_{p_i\in\mathbb{R},~i=1,\dots,K}~&\sum_{i=1}^K\alpha_i\log_2(1+\zeta_i(\{p_k\}_{k{\ne}i})Q_{ii}p_i)\\
        {\rm s.t.}~&0~ {\le}~ p_i~ {\le}~ P_i,~i=1,\dots,K,
        \end{align}
    \end{subequations}
where $\zeta_i(\{p_k\}_{k{\ne}i})$, which is continuously
differentiable in $\{p_k\}_{k{\ne}i}$, is the unique solution that
satisfies
\begin{align}\label{psi}
\Psi_i(x,\{p_k\}_{k{\ne}i})\!\triangleq\!\rho_i\exp(\sigma_i^2x)\prod_{k{\ne}i}\left(1+Q_{ki}p_k\cdot{x}\right)=1,
\end{align}
for all $i,k=1,\dots,K$.
\end{Lemma}

\emph{Proof:} See Appendix \ref{sec:proof of lemma_equivalent pwc}
for details.\hfill{$\blacksquare$}

By reformulating problem \eqref{WSRM_pwr} as problem
\eqref{WSRM_pwr_ctrl}, one can compactly characterize the relation
between the achievable rates and the transmit powers, e.g., the
monotonicity and convexity, using the implicit functions
$\zeta_i(\cdot)$, $i=1,\dots,K$. As a result, it is much easier to
analyze the optimal power allocation pattern based on the
alternative formulation \eqref{WSRM_pwr_ctrl} than based on the
original formulation \eqref{WSRM_pwr}. In the subsequent analysis,
we thus focus on proving that the Max-Cut problem is polynomial-time
reducible to problem \eqref{WSRM_pwr_ctrl}. The idea of this proof
is that, given any undirected and connected graph
$G=(\mathcal{V},\mathcal{E})$ and the weights $w_{ij}>0$ for all
$(i,j)\in\mathcal{E}$, we can construct a particular instance of
\eqref{WSRM_pwr_ctrl} that is equivalent to the weighted Max-Cut
problem \eqref{MAXCUT} associated with the graph $G$ and the weights
$\{w_{ij}\}$. The construction is detailed as follows.

We associate each node $i\in\mathcal{V}$ with two distinct
transmitter-receiver pairs (users) in the coordinated power control
problem \eqref{WSRM_pwr_ctrl}, denoted by $v_{i0},v_{i1}$. Moreover,
each edge $(i,j)\in\mathcal{E}$ is associated with two other users,
denoted by $e_{ij},e_{ji}$. The resulting set of users is the union
of the user set $\mathcal{U}_v$ associated with nodes and the user
set $\mathcal{U}_e$ associated with edges, i.e.,
\begin{align}
\mathcal{U}&=\mathcal{U}_v\cup\mathcal{U}_e\notag\\
&\triangleq\{v_{10},v_{11},\dots,v_{|\mathcal{V}|0},v_{|\mathcal{V}|1}\}\cup\{e_{ij},e_{ji}|~(i,j)\in\mathcal{E}\},\label{maxcut_ifc_usr_set}
\end{align}
which contains $K=|\mathcal{U}|=2(|\mathcal{V}|+|\mathcal{E}|)$
users in total. For these $K$ users, we consider a particular
instance of problem \eqref{WSRM_pwr_ctrl} with the following
specified system parameters:
\begin{subequations}\label{parameter_setting}
\begin{align}
&\sigma_u^2\!=\!0.1,~\rho_u\!=\!0.95,~P_u\!=\!\left\{\!\!\!\begin{array}{ll}
                                                   1,&\text{if}~u\in\mathcal{U}_v,\\
                                                   0.7,&\text{if}~u\in\mathcal{U}_e,
                                                   \end{array}\right.\!\forall{u}\!\in\!\mathcal{U},\label{parameter_setting_a}\\
&Q_{v_{ia}v_{jb}}\!=\!\left\{\!\!\!\begin{array}{ll}
                         1,&\text{if}~i=j,\\
                         0,&\text{otherwise,}
                         \end{array}\right.
\!\!Q_{e_{ij}u}\!=\!\left\{\!\!\!\begin{array}{ll}
                   1,&\text{if}~u=e_{ij},\\
                   0,&\text{otherwise},
                   \end{array}\right.\label{parameter_setting_b}\\
&Q_{v_{ia}e_{k\ell}}\!=\!\left\{\begin{array}{ll}
                            1,&\text{if}~(i,a)=(k,0)~\text{or}~(i,a)=(\ell,1),\\
                            0,&\text{otherwise},
                            \end{array}\right.\label{parameter_setting_c}\\
&\begin{cases}\alpha_{v_{ia}}\!=\!1,& \forall{v_{ia}}\in\mathcal{U}_v,\\
              \alpha_{e_{ij}}\!=\!\alpha_{e_{ji}}\!=\!\dfrac{\wb_{ij}}{2\sum_{(k,\ell)\in\mathcal{E}}w_{k,\ell}},& \forall{e}_{ij}\in\mathcal{U}_e,~i<j.
 \end{cases}\label{parameter_setting_d}
\end{align}
\end{subequations}

\begin{figure}[t]
\begin{center}
\psfrag{v1}[BC][BC]{\scriptsize $1$} \psfrag{v2}[BC][BC]{\scriptsize
$2$} \psfrag{v3}[BC][BC]{\scriptsize $3$}
\psfrag{12}[BC][BC]{\scriptsize $(1,2)$}
\psfrag{23}[BC][BC]{\scriptsize $(2,3)$}
\psfrag{v10}[BC][BC]{$v_{10}$} \psfrag{v11}[BC][BC]{$v_{11}$}
\psfrag{v20}[BC][BC]{$v_{20}$} \psfrag{v21}[BC][BC]{$v_{21}$}
\psfrag{v30}[BC][BC]{$v_{30}$} \psfrag{v31}[BC][BC]{$v_{31}$}
\psfrag{e12}[BC][BC]{$e_{12}$} \psfrag{e21}[BC][BC]{$e_{21}$}
\psfrag{e23}[BC][BC]{$e_{23}$} \psfrag{e32}[BC][BC]{$e_{32}$}
\psfrag{Tx}[BC][BC]{\small$\mathrm{Transmitters}$}
\psfrag{Rx}[BC][BC]{\small$\mathrm{Receivers}$}
\psfrag{DL}[BC][BC]{\small$\mathrm{Direct~Links}$}
\psfrag{CL}[BC][BC]{\small$\mathrm{Cross~Links}$}
\includegraphics[scale=0.7]{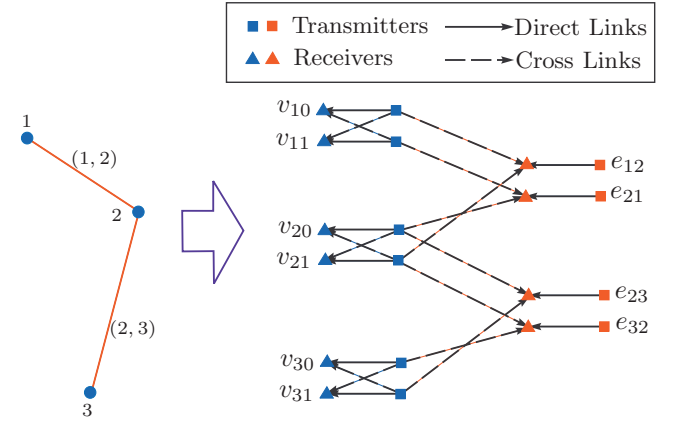}
\end{center}\vspace{-0.4cm}
\caption{An example illustrating the association between a graph
$G(\mathcal{V}\!=\!\{1,2,3\},~\mathcal{E}\!=\!\{(1,2),(2,3)\}\!)$
and an IFC with 10 users.}\vspace{-.5cm}\label{fig:G_to_IFC}
\end{figure}

To clarify the association described by \eqref{maxcut_ifc_usr_set}
and \eqref{parameter_setting}, let us consider a simple example
illustrated by Fig. \ref{fig:G_to_IFC}. Here, we demonstrate how a
simple graph $G$ with three vertices $\mathcal{V}=\{1,2,3\}$ and two
edges $\mathcal{E}=\{(1,2),(2,3)\}$ can be mapped to an IFC through
\eqref{maxcut_ifc_usr_set} and \eqref{parameter_setting}. As
illustrated in Fig. \ref{fig:G_to_IFC}, each vertex $i$ corresponds
to two users, $v_{i0}$ and $v_{i1}$, for all $i\in\{1,2,3\}$; and
each edge $(i,j)$ also corresponds to two users, $e_{ij}$ and
$e_{ji}$, for all $(i,j) \in \{(1,2),(2,3)\}$. For the IFC on the
right-hand side in Fig. \ref{fig:G_to_IFC}, user $u$ and user $u'$
would interfere with each other if $Q_{uu'}=1$ and do not interfere
if $Q_{uu'}=0$. Thus, according to the parameter set in
\eqref{parameter_setting_b}, any two users associated with a common
vertex in $G$ will interfere with each other, while users associated
with different vertices would not interfere with each other.
Besides, the users associated with the edges only communicate with
their intended receivers and do not interfere with the other users.
By \eqref{parameter_setting_c}, user $v_{i0}$ and $v_{j1}$ would
interfere with user $e_{ij}$, while users $v_{j0}$ and $v_{i1}$
would interfere with user $e_{ji}$, for all edges $(i,j)$. All the
resulting interference patterns are shown in Fig.
\ref{fig:G_to_IFC}.

Based on the construction described by \eqref{maxcut_ifc_usr_set}
and \eqref{parameter_setting}, the resulting problem instance of the
coordinated power control problem \eqref{WSRM_pwr_ctrl} is given by
\begin{subequations}\label{WSRM_SISO_graph}
\begin{align}
\max_{\substack{p_{ia}^v\in\mathbb{R},~\forall{v_{ia}}\in\mathcal{U}_v,\\p_{jk}^e\in\mathbb{R},~\forall{e_{jk}}\in\mathcal{U}_e}}&\sum_{v_{ia}\in\mathcal{U}_v}\!\!\log_2\!\big(1+p_{ia}^v\zeta^v(p_{i\bar{a}}^v)\big)\notag\\
&~~~+\!\!\sum_{e_{ij}\in\mathcal{U}_e}\!\!\alpha_{e_{ij}}\!\log_2\!\big(1\!+\!p_{ij}^e\zeta^e(p_{i0}^v,p_{j1}^v)\big)\label{WSRM_SISO_graph_a}\\
\text{s.t.}&~0\le{p}_{ia}^v\le1,~v_{ia}\in\mathcal{U}_v,\label{WSRM_SISO_graph_b}\\
&~0\le{p}_{ij}^e\le0.7,~e_{ij}\in\mathcal{U}_e,\label{WSRM_SISO_graph_c}
\end{align}
\end{subequations}
where $p_{ia}^v$ and $p_{jk}^e$ denote the transmission powers of
user $v_{ia}$ and $e_{jk}$, respectively; $\bar{a}=1-a$ for $a\in
\{0,1\}$; and by \eqref{psi}, $\zeta^v(p)$ and
$\zeta^e(p_1,p_2)=\zeta^e(p_2,p_1)$ are the unique solutions
of\vspace{-.3cm}
\begin{subequations}\label{def:zeta}
\begin{align}
\Psi^v(x,p)\triangleq&~0.95\cdot{e}^{0.1x}(1+px)=1,\label{def:zeta_v}\\
\Psi_e(x,p_1,p_2)\triangleq&~0.95\cdot{e}^{0.1x}(1+p_1x)(1+p_2x)=1,\label{def:zeta_e}
\end{align}
\end{subequations}
respectively. Next, we show that problem \eqref{WSRM_SISO_graph} is
equivalent to the Max-Cut problem \eqref{MAXCUT}. To this end, we
need to demonstrate that the optimal solution to problem
\eqref{WSRM_SISO_graph} lies in a discrete set, as stated in the
following lemma.

\vspace{-.3cm}

\begin{Lemma}\label{lemma:WSRM_SISO_graph_opt}
The optimal solution of \eqref{WSRM_SISO_graph} must satisfy
\begin{subequations}\label{Opt_pv}
\begin{align}
&\left[\begin{array}{c}p_{i0}^v\\p_{i1}^v\end{array}\right]\in\left\{\left[\begin{array}{c}0\\1\end{array}\right],
\left[\begin{array}{c}1\\0\end{array}\right]\right\}~\text{for}~i=1,\dots,|\mathcal{V}|,\\
&p_{jk}^e=0.7,~\forall{e_{jk}}\in\mathcal{U}_e.
\end{align}
\end{subequations}
\end{Lemma}

\emph{Proof:} Suppose that
$\{\hat{p}_{ia}^v,\hat{p}_{jk}^e\}_{v_{ia}\in\mathcal{U}_v,e_{jk}\in\mathcal{U}_e}$
is an optimal solution of problem \eqref{WSRM_SISO_graph}. It is
easy to see from \eqref{WSRM_SISO_graph_a} that $\hat{p}_{ij}^e=0.7$
for all $e_{ij}\in\mathcal{U}_e$, since user $e_{ij}$ does not
interfere with any other user.

To complete the proof, we first show that $\hat{p}_{ia}^v\in\{0,1\}$
for all $v_{ia}\in\mathcal{U}_v$. For notational simplicity, let us
focus on proving the case of $a=0$. For any given $i\in\mathcal{V}$,
let us assume that $\{\hat{p}_{jb}^v\}_{(j,b)\ne(i,0)}$ are known
and fixed. In this case, it is clear that $\hat{p}_{i0}^v$ must also
be optimal to the following problem:
\begin{subequations}\label{WSRM_SISO_one_user}
\begin{align}
\max_{p}~&F(p)\triangleq\log_2(1+p\zeta^v(\hat{p}_{i1}^v))+\log_2(1+\hat{p}_{i1}^v\zeta^v(p))\notag\\
&~~~~~~~~~+\!\!\sum_{\{j|e_{ij}\in\mathcal{U}_e\}}\!\!\alpha_{e_{ij}}\!\log_2(1\!+\!0.7\zeta^e(p,\hat{p}_{j1}^v))\label{WSRM_SISO_one_user_a}\\
\text{s.t.}~&0\le{p}\le1,\label{WSRM_SISO_one_user_b}
\end{align}
\end{subequations}
which is obtained from \eqref{WSRM_SISO_graph} by fixing
$p_{jb}^v=\hat{p}_{jb}^v$ for all $(j,b)\ne(i,0)$ and excluding the
terms irrelevant to user $v_{i0}$. Hence, we focus on showing that
the optimal solution to \eqref{WSRM_SISO_one_user} is either $p=0$
or $p=1$. To this end, we need the following lemma, which is proved
in Appendix \ref{sec:proof_lemma_quasiconvexity}.

\vspace{-.3cm}

\begin{Lemma}\label{lemma:quasiconvexity}
The objective function of problem \eqref{WSRM_SISO_one_user}, i.e.,
$F(p):~\mathbb{R}_+\to\mathbb{R}_+$, is a differentiable quasiconvex
function. Furthermore, there exists at most one $p\ge0$ such that
$dF(p)/dp=0$.
\end{Lemma}

Suppose that $p^\star\in(0,1)$ is optimal to
\eqref{WSRM_SISO_one_user}, which implies $dF(p^\star)/dp=0$. Then,
by Lemma \ref{lemma:quasiconvexity}, we have $dF(p)/dp<0$ for all
$p\in[0,p^\star)$ and $dF(p)/dp>0$ for all $p\in(p^\star,+\infty)$,
which contradicts the fact that $p^\star$ maximizes $F(p)$ over the
interval $0\le{p}\le1$, implying that the optimal solution to
problem \eqref{WSRM_SISO_one_user} is either $p=0$ or $p=1$. Hence,
we have proved $\hat{p}_{i0}^v\in\{0,1\}$. Similarly, we can show
that $\hat{p}_{i1}^v\in\{0,1\}$. As a result, the optimal solution
of problem \eqref{WSRM_SISO_graph} must satisfy
\[
\left[\begin{array}{c}\hat{p}_{i0}^v\\\hat{p}_{i1}^v\end{array}\right]
\in\left\{\left[\begin{array}{c}0\\1\end{array}\right],\left[\begin{array}{c}1\\0\end{array}\right],\left[\begin{array}{c}0\\0\end{array}\right],\left[\begin{array}{c}1\\1\end{array}\right]\right\},~i=1,\dots,|\mathcal{V}|,
\]
and $\hat{p}_{jk}^e=0.7$ for all $e_{jk}\in\mathcal{U}_e$.

What remains to be proved is that $[p_{i0}^v~p_{i1}^v]^T=[0~0]^T$ or
$[p_{i0}^v~p_{i1}^v]^T=[1~1]^T$ for some $i\in\mathcal{V}$ is
strictly suboptimal to problem \eqref{WSRM_SISO_graph}. Let
$\hat{R}$ denote the objective value of problem
\eqref{WSRM_SISO_graph} achieved by
$\{\hat{p}_{ia}^v,\hat{p}_{jk}^e\}_{v_{ia}\in\mathcal{U}_v,e_{jk}\in\mathcal{U}_e}$,
and define
\begin{align*}
&\mathcal{I}_0=\{i\in\mathcal{V}\mid[\hat{p}_{i0}^v~\hat{p}_{i1}^v]^T=[0~0]^T\},\\
&\mathcal{I}_1=\{i\in\mathcal{V}\mid[\hat{p}_{i0}^v~\hat{p}_{i1}^v]^T=[1~1]^T\}.
\end{align*}
We can bound $\hat{R}$ from below and above as follows:
\begin{subequations}\label{WSR_up_low_bounds}
\begin{align}
\hat{R}&\ge\sum_{v_{ia}\in\mathcal{U}_v}\log_2(1+\hat{p}_{ia}^v\zeta^v(\hat{p}_{i\bar{a}}^v))\notag\\
&=\sum_{i=1}^{|\mathcal{V}|}\bigg(\log_2(1+\hat{p}_{i0}^v\zeta^v(\hat{p}_{i1}^v))+\log_2(1+\hat{p}_{i1}^v\zeta^v(\hat{p}_{i0}^v))\bigg)\notag\\
&=(|\mathcal{V}|-|\mathcal{I}_0|-|\mathcal{I}_1|)\times\log_2(1\!+\!1\cdot\zeta^v(0))\notag\\
&~~~~+2|\mathcal{I}_1|\log_2(1+1\cdot\zeta^v(1))\notag\\
&=\left(|\mathcal{V}|-|\mathcal{I}_0|-|\mathcal{I}_1|\right)\times0.5973+2|\mathcal{I}_1|\times0.0671\notag\\
&=|\mathcal{V}|\times0.5973-(|\mathcal{I}_0|\times0.5973+|\mathcal{I}_1|\times0.4631)\label{WSR_up_low_bounds_a},\\
\hat{R}&\le\sum_{v_{ia}\in\mathcal{U}_v}\log_2(1+\hat{p}_{ia}^v\zeta^v(\hat{p}_{i\bar{a}}^v))\notag\\
&~~~~~~+\sum_{e_{jk}\in\mathcal{U}_e}\alpha_{e_{jk}}\log_2(1+0.7\cdot\zeta^e(0,0))~~\text{(by Lemma \ref{lemma:zeta_monotonicity})}\notag\\
&=|\mathcal{V}|\times0.5973-(|\mathcal{I}_0|\times0.5973+|\mathcal{I}_1|\times0.4631)\notag\\
&~~~~~~+\log_2(1+0.7\cdot\zeta^e(0,0))~~~~~~~~~~~~~~~~~~~~\text{(by \eqref{parameter_setting_b})}\notag\\
&=|\mathcal{V}|\times0.5973\notag\\
&~~~~~~-(|\mathcal{I}_0|\times0.5973+|\mathcal{I}_1|\times0.4631-0.4426),\label{WSR_up_low_bounds_b}
\end{align}
\end{subequations}
where the above numerical values are computed from the parameters
set in \eqref{parameter_setting}. From \eqref{WSR_up_low_bounds_a},
one can see that $\hat{R}$ is no less than
$|\mathcal{V}|\times0.5973$ when
$|\mathcal{I}_0|=|\mathcal{I}_1|=0$. However, from
\eqref{WSR_up_low_bounds_b}, it is clear that $\hat{R}$ is smaller
than $|\mathcal{V}|\times0.5973$, when either $|\mathcal{I}_0|\ne0$
or $|\mathcal{I}_1|\ne0$. Thus, the optimal power pattern
$[\hat{p}_{i0}^v~\hat{p}_{i1}^v]^T$ must be either $[0~1]^T$ or
$[1~0]^T$. This completes the proof of Lemma
\ref{lemma:WSRM_SISO_graph_opt}. \hfill{$\blacksquare$}

\vspace{.3cm}

Based on Lemma \ref{lemma:WSRM_SISO_graph_opt}, we can restrict the
feasible set of \eqref{WSRM_SISO_graph} to the subset defined in
\eqref{Opt_pv} without loss of optimality. Let
\[
\mathcal{S}=\left\{i\left|~[p_{i0}^v~p_{i1}^v]^T=[0~1]^T\right.\right\}\subseteq\mathcal{V},
\]
and denote $1_{\{i\in\mathcal{S}\}}$ as the indicator function which
is equal to one if $i\in\mathcal{S}$ and zero otherwise. Then, by
\eqref{Opt_pv}, we have $p_{i0}^v=1_{\{i\in\bar{\mathcal{S}}\}}$,
$p_{i1}^v=1_{\{i\in\mathcal{S}\}}$ and $p_{jk}^e=0.7$. With these
substitutions in problem \eqref{WSRM_SISO_graph}, problem
\eqref{WSRM_SISO_graph} can be equivalently reformulated as the
following problem where $\mathcal{S}$ is now the optimization
variable:
{\allowdisplaybreaks[4]\begin{align}
&\max_{\mathcal{S}\subseteq\mathcal{V}}~\sum_{i=1}^{|\mathcal{V}|}\left[\log_2(1+1_{\{i\in\bar{\mathcal{S}}\}}\zeta^v(1_{\{i\in\mathcal{S}\}}))\right.\notag\\
&~~~~~~~~~~~~~~+\left.\log_2(1+1_{\{i\in\mathcal{S}\}}\zeta^v(1_{\{i\in\bar{\mathcal{S}}\}}))\right]\notag\\
&~~~~~~~+\!\!\sum_{(i,j)\in\mathcal{E}}\!\!\left[\alpha_{e_{ij}}\!\log_2(1\!+\!0.7\zeta^e(1_{\{i\in\bar{\mathcal{S}}\}},1_{\{j\in\mathcal{S}\}}))\right.\notag\\
&~~~~~~~~~~~~~~~~~\left.+\alpha_{e_{ji}}\!\log_2\!\big(\!1\!+\!0.7\zeta^e(1_{\{j\in\bar{\mathcal{S}}\}},1_{\{i\in\mathcal{S}\}})\!\big)\!\right]\label{WSRM_MAXCUT_equiv_1}\\
&=\max_{\mathcal{S}\subseteq\mathcal{V}}~\sum_{i=1}^{|\mathcal{V}|}\log_2(1\!+\!1\cdot\zeta^v(0))+\!\!\sum_{(i,j)\in\mathcal{E}\backslash\mathcal{E}(\mathcal{S})}\!\!\alpha_{e_{ij}}(c_{01}\!+\!c_{10})\notag\\
&~~~~~~~+\sum_{(i,j)\in\mathcal{E}(\mathcal{S})}\!\!\alpha_{e_{ij}}(c_{00}\!+\!c_{11})\label{WSRM_MAXCUT_equiv_2}\\
&=\max_{\mathcal{S}\subseteq\mathcal{V}}~\sum_{i=1}^{|\mathcal{V}|}\log_2(1+1\cdot\zeta^v(0))+\sum_{(i,j)\in\mathcal{E}}\alpha_{e_{ij}}(c_{01}+c_{10})\notag\\
&~~~~~~~+\sum_{(i,j)\in\mathcal{E}(\mathcal{S})}\alpha_{e_{ij}}(c_{00}+c_{11}-c_{01}-c_{10})\label{WSRM_MAXCUT_equiv_3}\\
&=\max_{\mathcal{S}\subseteq\mathcal{V}}~|\mathcal{V}|\cdot\log_2(1+1\cdot\zeta^v(0))+\frac{c_{01}+c_{10}}{2}\notag\\
&~~~~~~~+\frac{c_{00}+c_{11}-c_{01}-c_{10}}{2\sum_{(i,j)\in\mathcal{E}}w_{ij}}\cdot\sum_{(i,j)\in\mathcal{E}(\mathcal{S})}w_{ij},\notag
\end{align}}
\hspace{-.1cm}where $c_{00}=\log_2(1+0.7\zeta^e(0,0))$,
$c_{11}=\log_2(1+0.7\zeta^e(1,1))$,
$c_{01}=\log_2(1+0.7\zeta^e(0,1))$, and
$c_{10}=\log_2(1+0.7\zeta^e(1,0))$ are constants. Note that, when
$(i,j)\in\mathcal{E}(\mathcal{S})$, vertex $i$ and vertex $j$ belong
to different subsets $\mathcal{S}$ and $\bar{\mathcal{S}}$, so it
holds true that
$1_{\{i\in\bar{\mathcal{S}}\}}=1_{\{j\in\mathcal{S}\}}$ and
$1_{\{i\in\mathcal{S}\}}=1_{\{j\in\bar{\mathcal{S}}\}}$; on the
other hand, when
$(i,j)\in\mathcal{E}\backslash\mathcal{E}(\mathcal{S})$, it holds
true that $1_{\{i\in\bar{\mathcal{S}}\}}\ne1_{\{j\in\mathcal{S}\}}$
and $1_{\{i\in\mathcal{S}\}}\ne1_{\{j\in\bar{\mathcal{S}}\}}$, by
which we obtain \eqref{WSRM_MAXCUT_equiv_2}. By \eqref{def:zeta_e},
one can show that $c_{00}+c_{11}-c_{01}-c_{10}>0$, so solving
\eqref{WSRM_SISO_graph} is equivalent to solving the Max-Cut problem
\eqref{MAXCUT}. Thus, we have presented a polynomial-time reduction
that converts all the problem instances of the Max-Cut problem
\eqref{MAXCUT} to instances of the outage constrained SRM CoBF
problem \eqref{WSRM}, which completes the proof of Theorem
\ref{thm:WSRM_complexity}. \hfill $\blacksquare$

\section{Complexity Analysis for Outage constrained MMF problem}\label{sec:WMRM_complexity_analysis_proof}

In this section, we turn our attention to the outage constrained MMF
problem, i.e.,
\begin{subequations}\label{WMRM}
\begin{align}
\max_{\substack{\wb_i\in\Cplx^{N_t},~R_i\ge0,\\i=1,\dots,K}}~&R\triangleq\min_{i=1,\dots,K}~R_i/\alpha_i\label{WMRM_a}\\
\text{s.t.}~&\rho_i\exp\left(\frac{(2^{R_i}-1)\sigma_i^2}{\wb_i^H\Qb_{ii}\wb_i}\right)\notag\\
&~\times\!\prod_{k{\ne}i}\!\left(\!1\!+\!\frac{(2^{R_i}\!-\!1)\wb_k^H\Qb_{ki}\wb_k}{\wb_i^H\Qb_{ii}\wb_i}\!\right)\!\le\!1,\label{WMRM_b}\\
&\|\wb_i\|^2{\le}P_i,~i=1,\dots,K.\label{WMRM_c}
\end{align}
\end{subequations}
In contrast to its perfect CSI counterpart, which can be transformed
into a quasiconvex problem for multiple antennas and multiple users
\cite{Liu_11}, we will show that problem \eqref{WMRM} is
polynomial-time solvable only for the single antenna case, i.e.,
$N_t=1$, but NP-hard in the number of users $K$ when $N_t\ge2$.

To proceed with the complexity analysis, let us first introduce a
feasibility problem. That is, given a target rate $\bar{R}\ge0$,
\begin{subequations}\label{WMRM_feasibility}
\begin{align}
\mathrm{Find}~&\wb_1,\dots,\wb_K\label{WMRM_feasibility_a}\\
\text{s.t.}~&\rho_i\exp\left(\frac{(2^{\alpha_i\bar{R}}-1)\sigma_i^2}{\wb_i^H\Qb_{ii}\wb_i}\right)\notag\\
&~\times\prod_{k{\ne}i}\left(1+\frac{(2^{\alpha_i\bar{R}}-1)\wb_k^H\Qb_{ki}\wb_k}{\wb_i^H\Qb_{ii}\wb_i}\right)\le1,\label{WMRM_feasibility_b}\\
&\|\wb_i\|^2 ~{\le}~P_i,~i=1,\dots,K.\label{WMRM_feasibility_c}
\end{align}
\end{subequations}
Note that problem \eqref{WMRM_feasibility} is closely related to
problem \eqref{WMRM}. Their relation is specified in the following
lemma.

\begin{Lemma}\label{lemma:WMRM_feasibility}
Let $R^\star$ denote the optimal value of problem \eqref{WMRM}. It
holds true that, for any $\bar{R}\ge0$, problem
\eqref{WMRM_feasibility} is feasible if and only if
$\bar{R}\le{R}^\star$. Furthermore, the set of optimal beamformers
to problem \eqref{WMRM} is a subset of the feasible set of problem
\eqref{WMRM_feasibility} when $\bar{R}<R^\star$, and these two sets
coincide when $\bar{R}=R^\star$.
\end{Lemma}

\emph{Proof:} Let $\{\wb_i^\star\}$ denote an optimal beamformer to
problem \eqref{WMRM}. Since the left-hand side of constraint
\eqref{WMRM_feasibility_b} is strictly increasing in $\bar{R}$, we
know that $\{\wb_i^\star\}$ is feasible to problem
\eqref{WMRM_feasibility} if $\bar{R}\le{R}^\star$. Hence, problem
\eqref{WMRM_feasibility} is feasible if $\bar{R}\le{R}^\star$, and
all the optimal beamformers to problem \eqref{WMRM} are feasible to
problem \eqref{WMRM_feasibility}. On the other hand, suppose that
problem \eqref{WMRM_feasibility} is feasible and $\{\bar{\wb}_i\}$
is a feasible point. Then, $\{\bar{\wb}_i\}$ is clearly a feasible
beamformer of \eqref{WMRM} that achieves objective value $\bar{R}$,
implying $R^\star\ge\bar{R}$. Furthermore, when $\bar{R}=R^\star$,
all the feasible beamformers to problem \eqref{WMRM_feasibility}
achieve the optimal objective value of problem \eqref{WMRM}, and
hence are optimal beamformers to \eqref{WMRM}.\hfill{$\blacksquare$}

Lemma \ref{lemma:WMRM_feasibility} infers that problem \eqref{WMRM}
is polynomial-time solvable if and only if problem
\eqref{WMRM_feasibility} is polynomial-time solvable. In particular
if problem \eqref{WMRM_feasibility} can be efficiently solved, then
problem \eqref{WMRM} can be efficiently solved by a bisection
algorithm which involves solving a series of problem
\eqref{WMRM_feasibility} \cite[\S 4.2.5]{BK:BoydV04}. On the other
hand, if one can solve problem \eqref{WMRM}, then one can correctly
determine whether problem \eqref{WMRM_feasibility} is feasible, and,
if yes, obtain a set of feasible beamformers. Therefore, these two
problems belong to the same complexity class. In the next two
subsections, we investigate the complexity of problem \eqref{WMRM}
for the SISO case ($N_t=1$), and MISO case ($N_t>1$) based on this
observation.

\subsection{Single-Antenna Case}\label{subsec:mmf Nt=1}

In this subsection, we consider the SISO case, and present a
polynomial-time bisection algorithm, which involves solving a finite
number of the feasibility problem \eqref{WMRM_feasibility}, for
solving problem \eqref{WMRM} to the global optimum. When $N_t=1$,
problem \eqref{WMRM} degenerates to the following power control
problem
\begin{subequations}\label{WMRM_Nt1}
\begin{align}
\max_{\substack{p_i\ge0,~R_i\ge0,\\i=1,\dots,K}}~&R\label{WMRM_Nt1_a}\\
\text{s.t.}~&\rho_i\exp\left(\frac{(2^{R_i}-1)\sigma_i^2}{p_iQ_{ii}}\right)\notag\\
&~\times\prod_{k{\ne}i}\left(1+\frac{(2^{R_i}-1)p_kQ_{ki}}{p_iQ_{ii}}\right)\le1,\label{WMRM_Nt1_b}\\
&p_i\le{P}_i,~i=1,\dots,K,\label{WMRM_Nt1_c}
\end{align}
\end{subequations}
where $p_i$ is the transmit power of transmitter $i$, and $Q_{ki}$
is the variance of the SISO channel $h_{ki}$, for all
$i,k=1,\dots,K$. Similarly, problem \eqref{WMRM_feasibility} reduces
to
\begin{subequations}\label{WMRM_Nt1_feasibility}
\begin{align}
\mathrm{Find}~&(p_1,\dots,p_K)\\
\text{s.t.}~&\rho_i\exp\left(\frac{(2^{\alpha_i\bar{R}}-1)\sigma_i^2}{p_iQ_{ii}}\right)\notag\\
&~\times\prod_{k{\ne}i}\left(1+\frac{(2^{\alpha_i\bar{R}}-1)p_kQ_{ki}}{p_iQ_{ii}}\right)\le1,\\
&p_i\le{P}_i,~i=1,\dots,K,
\end{align}
\end{subequations}
which is feasible if and only if the optimal value of
\eqref{WMRM_Nt1} is no less than $\bar{R}$. First of all, we show
that problem \eqref{WMRM_Nt1_feasibility} is polynomial-time
solvable.

Consider the change of variables $\tilde{p}_i=\ln{p_i}$, for
$i=1,\dots,K$. Then, we can solve problem
\eqref{WMRM_Nt1_feasibility} by solving the following convex problem
\begin{subequations}\label{WMRM_Nt1_feasibility_GP}
\begin{align}
\min_{\substack{\tilde{\alpha},~\tilde{p}_i,\\i=1,\dots,K}}~&\tilde{\alpha}\\
\text{s.t.}~&\ln\rho_i+\frac{(2^{\alpha_i\bar{R}}-1)\sigma_i^2}{Q_{ii}}{\cdot}e^{-\tilde{p}_i}\notag\\
&~+\!\sum_{k{\ne}i}\ln\!\left(\!1\!+\!\frac{(2^{\alpha_i\bar{R}}\!-\!1)Q_{ki}}{Q_{ii}}{\cdot}e^{-\tilde{p}_i+\tilde{p}_k}\!\right)\!\le\!\tilde{\alpha},\\
&\tilde{p}_i\le\ln{P_i},~i=1,\dots,K.
\end{align}
\end{subequations}
Specifically, it is not difficult to verify by using an argument
similar to Lemma \ref{lemma:WMRM_feasibility} that problem
\eqref{WMRM_Nt1_feasibility} is feasible if and only if the optimal
$\tilde{\alpha}$ of problem \eqref{WMRM_Nt1_feasibility_GP} is less
than or equal to zero, and that every optimal solution of problem
\eqref{WMRM_Nt1_feasibility_GP} directly serves as a feasible point
of problem \eqref{WMRM_Nt1_feasibility} provided that problem
\eqref{WMRM_Nt1_feasibility} is feasible. Therefore, based on Lemma
\ref{lemma:WMRM_feasibility}, one can solve problem \eqref{WMRM_Nt1}
in a bisection manner by solving a sequence of convex problem
\eqref{WMRM_Nt1_feasibility_GP}. The bisection algorithm is
described in Algorithm \ref{alg:bisection}.\footnote{The initial
bisection interval of $\bar R$ can be found as follows. Firstly,
$\bar R=0$ is obviously a lower bound to the optimal value of
\eqref{WMRM_Nt1}. Secondly, by \eqref{WMRM_Nt1_b}, we have
$\rho_i\exp\left(\frac{(2^{R_i}-1)\sigma_i^2}{p_iQ_{ii}}\right)\le1\label{WMRM_upbound_1},~i=1,\dots,K,$
and thus the optimal value of problem \eqref{WMRM_Nt1} is upper
bounded by $\bar R =\min_i\frac{1}{\alpha_i}\log_2\left(1+\frac{P_i
Q_{ii}\ln(1/\rho_i)}{\sigma_i^2}\right)$.}

\begin{algorithm}[h]
\caption{Bisection algorithm for solving problem
\eqref{WMRM_Nt1}}\label{alg:bisection}
\begin{algorithmic}[1]
\STATE {\bf Set} $\bar{R}_\ell:=0$,
$\bar{R}_u:=\min_i\frac{1}{\alpha_i}\log_2\left(1+\frac{P_iQ_{ii}\ln(1/\rho_i)}{\sigma_i^2}\right)$,
and set the solution accuracy to $\delta>0$;

\REPEAT

\STATE Set $\bar{R}:=(\bar{R}_\ell+\bar{R}_u)/2$;

\STATE Solve problem \eqref{WMRM_Nt1_feasibility_GP}, and denote the
solution as
$(\!\{\tilde{p}_i^\star\}_{i=1}^K,\tilde{\alpha}^\star\!)$;

\STATE Set $\bar{R}_\ell:=\bar{R}$ if $\tilde{\alpha}^\star\le0$;
otherwise, set $\bar{R}_u:=\bar{R}$;

\UNTIL $\bar{R}_u-\bar{R}_\ell<\delta$;

\STATE {\bf Output}
  $p_i=e^{\tilde{p}_i^\star}$, $R=\bar{R}$, $i=1,\dots,K$, as a
  solution to problem \eqref{WMRM_Nt1}.
\end{algorithmic}
\end{algorithm}

Problem \eqref{WMRM_Nt1_feasibility_GP} is in fact equivalent to the
outage balancing power control problem studied in \cite{Tan_2011},
which can be efficiently solved by a nonlinear Perron-Frobenius
theory-based algorithm with overall complexity of
$\mathcal{O}(K\ln(K/\epsilon))$ (see \cite[Algorithm 1]{Tan_2011}),
where $\epsilon$ specifies the solution accuracy. Therefore, the
overall complexity of Algorithm \ref{alg:bisection} is
$\kappa\cdot\mathcal{O}(K\ln(K/\epsilon))$, where
\[
\kappa\triangleq\left\lceil\log_2\left(\delta^{-1}\bar{R}_u\right)\right\rceil
\]
is the number of bisection iterations required for Algorithm
\ref{alg:bisection} to achieve a solution accuracy $\delta$.
Algorithm \ref{alg:bisection} has a polynomial-time complexity since
$\kappa$ is finite in practical situations. Hence, we have proven
the complexity of the MMF CoBF problem for the case of $N_t=1$ as
stated in the following theorem.

\begin{Theorem}\label{thm:SISO}
When $N_t=1$, the outage constrained MMF CoBF problem \eqref{WMRM}
(i.e., problem \eqref{WMRM_Nt1}) is polynomial-time solvable.
\end{Theorem}

\subsection{Multiple-Antenna Case}\label{subsec:mmf Nt>1}

In contrast to the single transmit antenna case, in this subsection, we show that the
outage constrained MMF CoBF problem \eqref{WMRM} is NP-hard when
each of the transmitters is equipped with multiple antennas.

\begin{Theorem}\label{thm:MISO}
When $N_t\ge2$, the outage constrained MMF CoBF problem \eqref{WMRM}
is NP-hard in the number of users $K$.
\end{Theorem}

\emph{Proof:} As inferred from Lemma \ref{lemma:WMRM_feasibility},
it suffices to show that solving the feasibility problem
\eqref{WMRM_feasibility} is NP-hard when $N_t\ge2$. The main idea of
the proof is to show that the 3-satisfiability (3-SAT) problem,
which is known to be NP-complete \cite{Karp2010_21NPComplete}, is
reducible to problem \eqref{WMRM_feasibility}. The 3-SAT problem is
defined as follows.

\begin{Definition}\label{Def:3-SAT}
Given $N$ Boolean variables and $M$ clauses each containing exactly
three literals of different Boolean variables, the 3-SAT problem is
to determine whether there exists a truth assignment of the Boolean
variables such that all the $M$ clauses hold true.
\end{Definition}

For ease of exposition, we use ``$\vee$'', ``$\neg$'' to denote the
logical disjunction (OR), negation, and formulate a 3-SAT problem as
a feasibility problem. Specifically, a 3-SAT problem instance with
$N$ Boolean variables $x_1,\dots,x_N$ and $M$ clauses
$d_m=y_{i_m}{\vee}y_{j_m}{\vee}y_{k_m}$,
$i_m,j_m,k_m\in\{1,\dots,N\}$, $m=1,\dots,M$, where $y_{i_m}$ is
either $x_{i_m}$ or its negation $\neg{x_{i_m}}$, and so are
$y_{j_m}$ and $y_{k_m}$, can be written as the following feasibility
problem:
\begin{subequations}\label{3-SAT}
\begin{align}
\mathrm{Find}~&[x_1,\dots,x_N]^T\in\{0,1\}^N\label{3-SAT_a}\\
\text{s.t.}~&d_m=y_{i_m}{\vee}y_{j_m}{\vee}y_{k_m}=1,~m=1,\dots,M.\label{3-SAT_b}
\end{align}
\end{subequations}
Given any $N$ Boolean variables $x_1,\dots,x_N$ and $M$ clauses
$d_1,\dots,d_M$, we construct a problem instance of
\eqref{WMRM_feasibility} that is equivalent to the corresponding
3-SAT problem, i.e., problem \eqref{3-SAT}, as follows.

We associate each $x_n$ with five users, denoted by the set
$\mathcal{V}_n\triangleq\{v_{n0},v_{n1},v_{n2},v_{n3},v_{n4}\}$, for
$n=1,\dots,N$, and associate the $M$ clauses $d_1,\dots,d_M$ with
$M$ users $\mathcal{C}\triangleq\{c_1,\dots,c_M\}$. The entire set
of users is thus
\[
\mathcal{U}=\mathcal{V}_1\cup\cdots\cup\mathcal{V}_N\cup\mathcal{C},
\]
which contains a total of $K=5N+M$ users. For these $K$ users, we
consider a particular problem instance of \eqref{WMRM_feasibility}
with the following specified system parameters:
\begin{subequations}\label{parameter_WMRM}
\begin{align}
&N_t\!=\!2,~\bar{R}\!=\!1,~\alpha_u\!=\!1,~P_u\!=\!1,~\rho_u\!=\!\rho\!=\!0.9,~{\forall}u\!\in\!\mathcal{U},\label{parameter_WMRM_a}\\
&\sigma_{v_{n\ell}}^2=\begin{cases}\ln\rho^{-1},&\ell=0,\\\ln(\frac{10}{11}\cdot\rho^{-1}),&\ell\ne0,\end{cases}~\sigma_{c_m}^2\!=\!0.01,~\forall{n,m},\label{parameter_WMRM_b}\\
&\Qb_{uu}\!=\!\begin{bmatrix}1&\!\!0\\0&\!\!1\end{bmatrix}\!,~{\forall}u\!\in\!\mathcal{U},~\Qb_{v_{n0},v_{n\ell}}\!=\!\frac{\Ab_\ell}{10},~\forall{n},\forall\ell\ne0,\label{parameter_WMRM_c}\\
&\Qb_{v_{n0},c_m}\!\!=\!\left\{\!\!\!\!\begin{array}{ll}
                         \frac{1}{25}\!\begin{bmatrix}0&\!\!0\\0&\!\!1\end{bmatrix}\!,&\!\!\text{if}~x_n\!\in\!\{y_{i_m},y_{j_m},y_{k_m}\},\\
                         \frac{1}{25}\!\begin{bmatrix}1&\!\!0\\0&\!\!0\end{bmatrix}\!,&\!\!\text{if}~\neg{x_n}\!\in\!\{y_{i_m},y_{j_m},y_{k_m}\},
                         \end{array}\right.\!\!\forall{n},\label{parameter_WMRM_d}\\
&\Qb_{u_1,u_2}=\zerob,~\text{for all $u_1$, $u_2$ not specified
above},\label{parameter_WMRM_e}
\end{align}
\end{subequations}
where
\[
\Ab_1\!=\!\begin{bmatrix}1&\!\!1\\1&\!\!1\end{bmatrix}\!,~\Ab_2\!=\!\begin{bmatrix}1&\!\!-1\\-1&\!\!1\end{bmatrix}\!,~\Ab_3\!=\!\begin{bmatrix}1&\!\!\iota\\-\iota&\!\!1\end{bmatrix}\!,~\Ab_4\!=\!\begin{bmatrix}1&\!\!-\iota\\\iota&\!\!1\end{bmatrix}\!,
\]
and $\iota^2=-1$.

\begin{figure}[t]
\begin{center}
\psfrag{clause1}[BC][BC]{\small$d_1=x_1{\vee}x_2{\vee}x_3$}
\psfrag{clause2}[BC][BC]{\small$d_2=x_2{\vee}\neg{x_3}{\vee}\neg{x_4}$}
\psfrag{c1}[BC][BC]{\small$c_1$} \psfrag{c2}[BC][BC]{\small$c_2$}
\psfrag{x1}[BC][BC]{\small$x_1$} \psfrag{x2}[BC][BC]{\small$x_2$}
\psfrag{x3}[BC][BC]{\small$x_3$} \psfrag{x4}[BC][BC]{\small$x_4$}
\psfrag{v10}[BC][BC]{\small$v_{10}$}
\psfrag{v11}[BC][BC]{\small$v_{11}$}
\psfrag{v12}[BC][BC]{\small$v_{12}$}
\psfrag{v13}[BC][BC]{\small$v_{13}$}
\psfrag{v14}[BC][BC]{\small$v_{14}$}
\psfrag{v20}[BC][BC]{\small$v_{20}$}
\psfrag{v21}[BC][BC]{\small$v_{21}$}
\psfrag{v22}[BC][BC]{\small$v_{22}$}
\psfrag{v23}[BC][BC]{\small$v_{23}$}
\psfrag{v24}[BC][BC]{\small$v_{24}$}
\psfrag{v30}[BC][BC]{\small$v_{30}$}
\psfrag{v31}[BC][BC]{\small$v_{31}$}
\psfrag{v32}[BC][BC]{\small$v_{32}$}
\psfrag{v33}[BC][BC]{\small$v_{33}$}
\psfrag{v34}[BC][BC]{\small$v_{34}$}
\psfrag{v40}[BC][BC]{\small$v_{40}$}
\psfrag{v41}[BC][BC]{\small$v_{41}$}
\psfrag{v42}[BC][BC]{\small$v_{42}$}
\psfrag{v43}[BC][BC]{\small$v_{43}$}
\psfrag{v44}[BC][BC]{\small$v_{44}$}
\psfrag{Tx}[BC][BC]{\small$\mathrm{Transmitters}$}
\psfrag{Rx}[BC][BC]{\small$\mathrm{Receivers}$}
\psfrag{DL}[BC][BC]{\small$\mathrm{Direct~Links}$}
\psfrag{CL}[BC][BC]{\small$\mathrm{Cross~Links}$}
\includegraphics[scale=0.67]{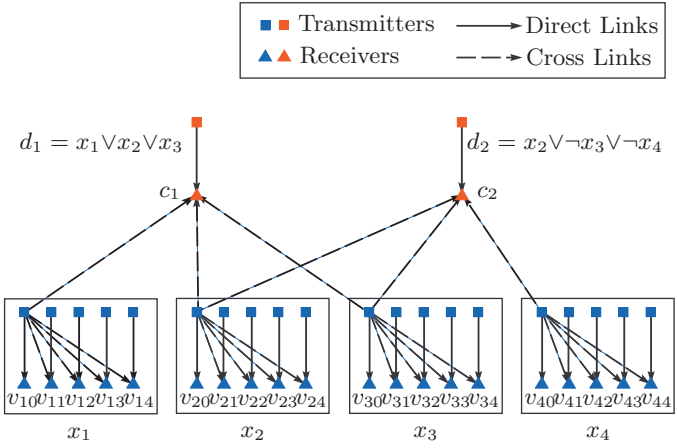}
\end{center}
\caption{An example illustrating the association between a 3-SAT
problem instance $d_1=(x_1{\vee}x_2{\vee}x_3)$,
$d_2=(x_2{\vee}\neg{x_3}{\vee}\neg{x_4})$ and an outage constrained
MMF CoBF problem \eqref{WMRM} with $N_t=2$ and $K=22$
users.}\label{fig:SAT_to_IFC}\vspace{-0.5cm}
\end{figure}

An illustrative example for the association between the 3-SAT
problem and the outage constrained MMF CoBF problem \eqref{WMRM}
described by \eqref{parameter_WMRM} is depicted in Fig.
\ref{fig:SAT_to_IFC}. In this example, we consider a 3-SAT problem
with four variables, $x_1,x_2,x_3,x_4$, and two clauses,
$d_1=y_{i_1}{\vee}y_{j_1}{\vee}y_{k_1}$,
$d_2=y_{i_2}{\vee}y_{j_2}{\vee}y_{k_2}$, where $y_{i_1}=x_1$,
$y_{j_1}=x_2$, $y_{k_1}=x_3$, $y_{i_2}=x_2$, $y_{j_2}=\neg{x}_3$,
and $y_{k_2}=\neg{x}_4$. In the corresponding IFC, each Boolean
variable $x_n$ is associated with five users, i.e.,
$\mathcal{V}_n=\{v_{n0},v_{n1},\dots,v_{n4}\}$, for $n=1,\dots,4$.
For each $n=1,\dots,4$, only user $v_{n0}$ interferes with users
$v_{n1},\dots,v_{n4}$ where the cross-link channel covariance
matrices are given in \eqref{parameter_WMRM_c}. Furthermore, user
$c_m$, which is due to clause $d_m$, is interfered by user $v_{n0}$
if the Boolean variable $x_n$ or its negation $\neg{x_n}$ appears in
clause $d_m$. As indicated by \eqref{parameter_WMRM_d}, the
covariance matrix of the channel from $v_{n0}$ to $c_m$ depends on
whether $x_n$ or the negation $\neg{x_n}$ appears in $d_m$. In Fig.
\ref{fig:SAT_to_IFC}, the users are not connected if the associated
cross-link channel covariance matrix is zero as indicated by
\eqref{parameter_WMRM_e}.

With the specified system parameters given in
\eqref{parameter_WMRM}, the constructed problem instance of
\eqref{WMRM_feasibility} can be expressed as
\begin{subequations}\label{WMRM_feasibility_instc}
\begin{align}
\mathrm{Find}~&\{\wb_{1\ell}^v\}_{\ell=0}^4,\dots,~\{\wb_{N\ell}^v\}_{\ell=0}^4,~\{\wb_m^c\}_{m=1}^M\label{WMRM_feasibility_instc_a}\\
\text{s.t.}~&\rho{e}^{\frac{\sigma_{v_{n0}}^2}{\|\wb_{n0}^v\|^2}}\le1,~\forall{n},\label{WMRM_feasibility_instc_b}\\
&\rho{e}^{\frac{\sigma_{v_{n\ell}}^2}{\|\wb_{n\ell}^v\|^2}}\!\bigg(\!1\!+\!\frac{(\wb_{n0}^v)^H\!\Ab_\ell\wb_{n0}^v}{10\|\wb_{n\ell}^v\|^2}\!\bigg)\!\!\le\!1,~\forall{n},~\forall\ell\ne0,\label{WMRM_feasibility_instc_c}\\
&\rho{e}^{\frac{\sigma_{c_m}^2}{\|\wb_m^c\|^2}}\!\!\!\!\prod_{\tau_m\in\{i_m,j_m,k_m\}}\!\!\!\bigg(1\!+\!\frac{(\wb_{\tau_m0}^v)^H\Qb_{v_{\tau_m0},c_m}\wb_{\tau_m0}^v}{\|\wb_m^c\|^2}\bigg)\notag\\
&~~~~~~~~~~~~~~~~~~~~~~~~~~~~~~~~~~~~~~~~~~~~\le1,~\forall{m},\label{WMRM_feasibility_instc_d}\\
&\|\wb_{n\ell}^v\|^2\le1,~\|\wb_m^c\|^2\le1,~\forall{n},~\forall{m},~\forall\ell,\label{WMRM_feasibility_instc_e}
\end{align}
\end{subequations}
where $\wb_{n\ell}^v$ and $\wb_m^c$ denote the beamformers of users
$v_{n\ell}$ and $c_m$, respectively. Next, we show that the given
3-SAT problem instance is satisfiable if and only if
\eqref{WMRM_feasibility_instc} is feasible.

To begin with, we show that any feasible point of problem
\eqref{WMRM_feasibility_instc} corresponds to a truth assignment
satisfying the 3-SAT problem \eqref{3-SAT}. Note that constraint
\eqref{WMRM_feasibility_instc_b} can be written as
\begin{equation}
\|\wb_{n0}^v\|^2\ge\sigma_{v_{n0}}^2(\ln\rho^{-1})^{-1}=1,~n=1,\dots,N,
\end{equation}
where the equality comes from \eqref{parameter_WMRM_b}. Combining
constraints \eqref{WMRM_feasibility_instc_b} with
\eqref{WMRM_feasibility_instc_e}, we have $\|\wb_{n0}^v\|^2=1$ for
$n=1,\dots,N$. Similarly, one can rewrite
\eqref{WMRM_feasibility_instc_c} as
\begin{align}
&(\wb_{n0}^v)^H\Ab_\ell\wb_{n0}^v\le10\|\wb_{n\ell}^v\|^2\left(\rho^{-1}
\exp\left(\frac{-\sigma_{v_{n\ell}}^2}{\|\wb_{n\ell}^v\|^2}\right)\!-\!1\right)\notag\\
&{\le}\max_{\|\wb_{n\ell}^v\|^2\le1}\!10\|\wb_{n\ell}^v\|^2\left(\rho^{-1}
\exp\left(\frac{-\sigma_{v_{n\ell}}^2}{\|\wb_{n\ell}^v\|^2}\right)\!-\!1\right)\!{=}1,\label{eq:power_1}
\end{align}
for $n=1,\dots,N$, $\ell=1,2,3,4$, where the 2nd inequality holds
with equality if and only if $\|\wb_{n\ell}^v\|^2=1$, and the last
equality comes from \eqref{parameter_WMRM_b}. Furthermore,
\begin{equation}
(\wb_{n0}^v)^H\Ab_\ell\wb_{n0}^v+(\wb_{n0}^v)^H\Ab_{\ell+1}\wb_{n0}^v=2\|\wb_{n0}^v\|^2=2,\label{eq:power_2}
\end{equation}
for $n=1,\dots,N$, $\ell=1,3$. Combining \eqref{eq:power_1} and
\eqref{eq:power_2}, we obtain
\begin{equation}\label{eq:power_constraint}
(\wb_{n0}^v)^H\Ab_\ell\wb_{n0}^v=1,~~\text{and}~~\|\wb_{n\ell}^v\|^2=1,
\end{equation}
for all $n=1,\dots,N$ and $\ell=1,2,3,4$. Then, we can derive
\begin{subequations}\label{eq:orthogonal_constraint}
\begin{align}
&(\wb_{n0}^v)^H\!(\Ab_1\!-\!\Ab_2)\wb_{n0}^v\!=\!4\mathrm{Re}\{([\wb_{n0}^v]_2)^H[\wb_{n0}^v]_1\}\!=\!0,\\
&(\wb_{n0}^v)^H\!(\Ab_3\!-\!\Ab_4)\wb_{n0}^v\!=\!4\mathrm{Im}\{([\wb_{n0}^v]_2)^H[\wb_{n0}^v]_1\}\!=\!0,
\end{align}
\end{subequations}
for all $n=1,\dots,N$, where $[\wb_{n0}^v]_1$ and $[\wb_{n0}^v]_2$
denote the first and second elements of $\wb_{n0}^v$, respectively.
Thus, a feasible $\wb_{n0}^v$ must satisfy either $[\wb_{n0}^v]_1=0$
or $[\wb_{n0}^v]_2=0$. In addition, constraints
\eqref{WMRM_feasibility_instc_d} and
\eqref{WMRM_feasibility_instc_e} imply that
\begin{align}
&\min_{\|\wb_m^c\|^2\le1}\rho{e}^{\frac{\sigma_{c_m}^2}{\|\wb_c^m\|^2}}\!\!\!\!\!\!\prod_{\tau_m\!\in\!\{i_m,j_m,k_m\}}\!\!\!\bigg(\!1\!+\!\frac{(\wb_{\tau_m0}^v)^H\!\Qb_{v_{\tau_m0},c_m}\wb_{\tau_m0}^v}{\|\wb_c^m\|^2}\!\bigg)\notag\\
&=\rho{e}^{\sigma_{c_m}^2}\!\!\!\!\!\prod_{\tau_m\!\in\!\{i_m,j_m,k_m\}}\!\!\!\!\big(\!1\!+\!(\wb_{\tau_m0}^v)^H\Qb_{v_{\tau_m0},c_m}\wb_{\tau_m0}^v\!\big)\!\le\!1.\label{eq:clause_equiv}
\end{align}
By combining \eqref{eq:power_constraint},
\eqref{eq:orthogonal_constraint}, and \eqref{eq:clause_equiv}, we
come up with the result that the feasible beamformers to problem
\eqref{WMRM_feasibility_instc} must satisfy
\begin{subequations}\label{WMRM_feasibility_instc_equiv}
\begin{align}
&\wb_{n0}^v\in\left\{\mathop{\cup}_{\theta\in[0,2\pi]}[e^{j\theta}~0]^T\right\}\cup\left\{\mathop{\cup}_{\theta\in[0,2\pi]}[0~e^{j\theta}]^T\right\},~\forall{n},\label{WMRM_feasibility_instc_equiv_b}\\
&\rho{e}^{\sigma_{c_m}^2}\!\!\!\prod_{\tau_m\in\{i_m,j_m,k_m\}}\!\!\big(1\!+\!(\wb_{\tau_m0}^v)^H\Qb_{v_{\tau_m0},c_m}\wb_{\tau_m0}^v\big)\!\le\!1,~\forall{m},\label{WMRM_feasibility_instc_equiv_c}\\
&\|\wb_{n\ell}^v\|^2=1,~\forall{n},~\forall{\ell}.\label{WMRM_feasibility_instc_equiv_d}
\end{align}
\end{subequations}
By \eqref{parameter_WMRM_d} and constraint
\eqref{WMRM_feasibility_instc_equiv_b}, one can see that
\begin{align}
&(\wb_{\tau_m0}^v)^H\Qb_{v_{\tau_m0},c_m}\wb_{\tau_m0}^v=\notag\\
&\begin{cases}
  0,&\text{if $y_{\tau_m}=x_{\tau_m}$,
  $\wb_{\tau_m0}^v\in\big\{{\displaystyle\mathop{\cup}_{\theta\in[0,2\pi]}}[e^{j\theta}~0]^T\big\}$},\\
  0,&\text{if $y_{\tau_m}=\neg{x}_{\tau_m}$,
  $\wb_{\tau_m0}^v\in\big\{{\displaystyle\mathop{\cup}_{\theta\in[0,2\pi]}}[0~e^{j\theta}]^T\big\}$},\\
  1/25,&\text{if $y_{\tau_m}=x_{\tau_m}$,
  $\wb_{\tau_m0}^v\in\big\{{\displaystyle\mathop{\cup}_{\theta\in[0,2\pi]}}[0~e^{j\theta}]^T\big\}$},\\
  1/25,&\text{if $y_{\tau_m}=\neg{x}_{\tau_m}$,
  $\wb_{\tau_m0}^v\in\big\{{\displaystyle\mathop{\cup}_{\theta\in[0,2\pi]}}[e^{j\theta}~0]^T\big\}$},
  \end{cases}\label{NP-hard_intrfr_powr}
\end{align}
where $\tau_m\in\{i_m,j_m,k_m\},~\forall{m}=1,\ldots,M$. Suppose
that $\hat{\wb}_{n\ell}^v$, $\hat{\wb}_m^c$, $n=1,\dots,N$,
$m=1,\dots,M$, $\ell=0,\dots,4$, are feasible to
\eqref{WMRM_feasibility_instc}, and set
\begin{equation}\label{eq:truth_assignment}
x_n=\begin{cases}
    1,&\text{if $\hat{\wb}_{n0}^v\in\big\{{\displaystyle\mathop{\cup}_{\theta\in[0,2\pi]}}[e^{j\theta}~0]^T\big\}$},\\
    0,&\text{if
    $\hat{\wb}_{n0}^v\in\big\{{\displaystyle\mathop{\cup}_{\theta\in[0,2\pi]}}[0~e^{j\theta}]^T\big\}$}.
    \end{cases}
\end{equation}
Then, we have
$(\hat{\wb}_{\tau_m0}^v)^H\Qb_{v_{\tau_m0},c_m}\hat{\wb}_{\tau_m0}^v=0$
if and only if $y_{\tau_m}=1$ according to
\eqref{NP-hard_intrfr_powr}. Furthermore, it can be verified by
\eqref{parameter_WMRM_a}, \eqref{parameter_WMRM_b} and
\eqref{NP-hard_intrfr_powr} that condition
\eqref{WMRM_feasibility_instc_equiv_c} is satisfied if and only if
$(\wb_{\tau_m0}^v)^H\Qb_{v_{\tau_m0},c_m}\wb_{\tau_m0}^v=0$, for
some $\tau_m\in\{i_m,j_m,k_m\}$, $\forall{m}$, implying that all of
the clauses $d_1,\dots,d_M$ are true. Hence, the truth assignment
\eqref{eq:truth_assignment} satisfies the given 3-SAT problem
instance \eqref{3-SAT}.

On the other hand, suppose that the given 3-SAT problem
\eqref{3-SAT} can be satisfied and $\hat{x}_n\in\{0,1\}$,
$n=1,\dots,N$, is a truth assignment such that the clauses
$d_1,\dots,d_M$ are true, i.e., constraints \eqref{3-SAT_b} are
satisfied. Then, analogous to the above analysis, it can be verified
that
\begin{align*}
&\wb_{n0}^v=\begin{cases}
            [1~0]^T,&\hat{x}_n=1,\\
            [0~1]^T,&\hat{x}_n=0,
            \end{cases}
 ~\forall{n},\\
&\|\wb_{n\ell}^v\|^2=\|\wb_m^c\|^2=1,~\forall{n},~\forall{m},~\forall\ell\ne0,
\end{align*}
is feasible to problem \eqref{WMRM_feasibility_instc}. Thus, we have
proven the equivalence of the given 3-SAT problem \eqref{3-SAT} and
problem \eqref{WMRM_feasibility_instc}, namely, the 3-SAT problem is
reducible to problem \eqref{WMRM_feasibility}. As a result,
determining the feasibility of problem \eqref{WMRM_feasibility} is
NP-hard when $N_t\ge2$, and hence solving problem \eqref{WMRM} is
also NP-hard according to Lemma
\ref{lemma:WMRM_feasibility}.\hfill{$\blacksquare$}

\begin{Remark}
The preceding proof of the NP-hardness of the feasibility problem
\eqref{WMRM_feasibility} can also be analogously applied to show the
NP-hardness of the following CoBF design problem for outage
balancing \cite{Tan_2011}:
\begin{subequations}\label{outage-balancing_CoBF}
\begin{align}
\min_{\substack{\wb_i\in\mathbb{C}^{N_t}\!,~\rho_i\in[0,1],\\i=1,\dots,K}}~&1/\rho\\
\text{s.t.}~&\rho_i\exp\!\left(\!\frac{(2^{R_i}-1)\sigma_i^2}{\wb_i^H\Qb_{ii}\wb_i}\!\right)\notag\\
&\times\!\prod_{k{\ne}i}\!\left(\!1\!+\!\frac{(2^{R_i}\!-\!1)\wb_k^H\Qb_{ki}\wb_k}{\wb_i^H\Qb_{ii}\wb_i}\!\right)\!\!\le\!1,\\
&\|\wb_i\|^2{\le}P_i,~i=1,\dots,K,
\end{align}
\end{subequations}
where $\rho=\min\{\rho_1,\ldots,\rho_K\}$ and $R_i$ is the given
data rate of the $i$th user, $i=1,\ldots,K$. Then, following a
similar argument as Lemma \ref{lemma:WMRM_feasibility}, one can show
that problem \eqref{outage-balancing_CoBF} is solvable (e.g., by
bisecting over $\rho\in [0,1]$) if and only if the feasibility
problem \eqref{WMRM_feasibility} (with $\rho_i=\rho$,
$\alpha_i\bar{R}=R_i$, $i=1,\dots,K$) is solvable. Therefore, the
NP-hardness of problem \eqref{WMRM_feasibility} when $N_t \geq 2$
also implies that problem \eqref{outage-balancing_CoBF} is NP-hard
when $N_t\ge2$.
\end{Remark}

\section{Conclusions}\label{sec:conclusion}

In this paper, we have studied the complexity status of the outage
constrained SRM and MMF CoBF problems. In particular, we have
established the NP-hardness of the two problems by showing that they
are at least as difficult as the Max-Cut problem and the 3-SAT
problem, respectively. Besides, a subclass, i.e., the SISO case, of
the outage constrained MMF problem is identified polynomial-time
solvable. Since the MMF CoBF problem is known polynomial-time
solvable under perfect CSI, our result implies that the outage
constrained CoBF design problems are indeed more challenging.
Motivated by our complexity analysis results, efficient algorithms
for obtaining high-quality approximate solutions to the outage
constrained CoBF problems are further pursued in the companion paper
\cite{Li_14_alg}.

\appendices {\setcounter{equation}{0}
\renewcommand{\theequation}{A.\arabic{equation}}

\section{Proof of Lemma \ref{lemma:equivalent pwc}}\label{sec:proof of lemma_equivalent pwc}

Notice that the function $\Psi_i(x,\{p_k\}_{k{\ne}i})$ in
\eqref{psi} is continuously differentiable with respect to
$(x,\{p_k\}_{k{\ne}i})$, and is strictly increasing in $x$. By the
implicit function theorem \cite{Krantz_Parks02}, there exists a
unique continuously differentiable function
$\zeta_i(\{p_k\}_{k{\ne}i})$ satisfying
\[
\Psi_i(\zeta_i(\{p_k\}_{k{\ne}i}),\{p_k\}_{k{\ne}i})=1,~\forall\{p_k\}_{k{\ne}i}.
\]
Therefore, we can equivalently express the rate outage constraint
\eqref{WSRM_pwr_b} as
\[
\frac{2^{R_i}\!-\!1}{Q_{ii}p_i}\!\le\!\zeta_i(\{p_k\}_{k{\ne}i})~\Leftrightarrow~R_i\!\le\!\log_2(1+\zeta_i(\{p_k\}_{k{\ne}i})Q_{ii}p_i),
\]
for $i=1,\dots,K$. Moreover, the objective function of problem
\eqref{WSRM_pwr} is nondecreasing with respect to $R_1,\dots,R_K$,
respectively. So, without loss of optimality, we can assume that
equality holds. Therefore, problem \eqref{WSRM_pwr} is equivalent to
problem \eqref{WSRM_pwr_ctrl}.\hfill{$\blacksquare$}

\section{Proof of Lemma \ref{lemma:quasiconvexity}}\label{sec:proof_lemma_quasiconvexity}

Let $f(p)$ denote the derivative of $F(p)$ with respect to $p$. It can be obtained as follows.
\begin{align*}
f(p)\ln2=&\frac{dF(p)}{dp}\cdot\ln2\\
=&\frac{\zeta^v(\hat{p}_{i1}^v)}{1+p\zeta^v(\hat{p}_{i1}^v)}+\frac{\hat{p}_{i1}^v}{1+\hat{p}_{i1}^v\zeta^v(p)}\frac{d\zeta^v(p)}{dp}\\
&+\sum_{\{j|e_{ij}\in\mathcal{U}_e\}}\frac{0.7\alpha_{e_{ij}}}{1+0.7\zeta^e(p,\hat{p}_{j1}^v)}\frac{d\zeta^e(p,\hat{p}_{j1}^v)}{dp}
\end{align*}
By applying implicit function theorem \cite{Krantz_Parks02} to
\eqref{def:zeta}, we can obtain closed-from expressions for the
derivatives of $\zeta^v(p)$ and $\zeta^e(p,\hat{p}_{j1}^v)$, which
are given in \eqref{zeta_v_deriv} and \eqref{zeta_e_deriv} on the
top of the next page. Hence, $f(p)\ln2$ can be expressed in
closed-from as \eqref{fpln2} on the top of the next page.

\begin{figure*}
\begin{subequations}
\begin{align}
&\frac{d\zeta^v(p)}{dp}=-\frac{\zeta^v(p)}{\sigma^2+\sigma^2p\zeta^v(p)+p}\label{zeta_v_deriv}\\
&\frac{d\zeta^e(p,\hat{p}_{j1}^v)}{dp}=-\frac{\zeta^e(p,\hat{p}_{j1}^v)(1+\hat{p}_{j1}^v\zeta^e(p,\hat{p}_{j1}^v))}{(1\!+\!p\zeta^e(p,\hat{p}_{j1}^v))[\hat{p}_{j1}^v\!+\!\sigma^2(1+\hat{p}_{j1}^v\zeta^e(p,\hat{p}_{j1}^v))]\!+\!p(1\!+\!\hat{p}_{j1}^v\zeta^e(p,\hat{p}_{j1}^v))}\label{zeta_e_deriv}\\
&f(p)\ln2=\frac{\zeta^v(\hat{p}_{i1}^v)}{1+p\zeta^v(\hat{p}_{i1}^v)}-\frac{\hat{p}_{i1}^v\zeta^v(p)}{1+\hat{p}_{i1}^v\zeta^v(p)}\frac{1}{\sigma^2+\sigma^2p\zeta^v(p)+p}\notag\\
&~~~~~~~~~~~~~~~~-\!\!\!\sum_{\{j|e_{ij}\in\mathcal{U}_e\}}\!\!\frac{\alpha_{e_{ij}}\cdot0.7\zeta^e(p,\hat{p}_{j1}^v)}{1+0.7\zeta^e(p,\hat{p}_{j1}^v)}\frac{1+\hat{p}_{j1}^v\zeta^e(p,\hat{p}_{j1}^v)}{(1+p\zeta^e(p,\hat{p}_{j1}^v))[\hat{p}_{j1}^v+\sigma^2(1+\hat{p}_{j1}^v\zeta^e(p,\hat{p}_{j1}^v))]+p(1+\hat{p}_{j1}^v\zeta^e(p,\hat{p}_{j1}^v))}\label{fpln2}
\end{align}
\end{subequations}
\hrulefill
\end{figure*}

Our goal is to show that, for any $\bar{P}\in[0,1]$, it holds true
that\vspace{-.2cm}
\begin{subequations}\label{derivative}
\begin{align}
&f(p)>0~\text{for
all}~p>\bar{P}~\text{if}~f(\bar{P})\ge0,\label{derivative_positive}\\
&f(p)<0~\text{for
all}~p<\bar{P}~\text{if}~f(\bar{P})\le0.\label{derivative_negative}
\end{align}
\end{subequations}
Since $f(p)$ is a continuous function, conditions in
\eqref{derivative} imply that either of the following two statements must
be true.
\begin{enumerate}
\item There exists $0\le\bar{P}\le1$
such that $f(\bar{P})=0$, which, according to
\eqref{derivative_positive} and \eqref{derivative_negative}, implies
that $f(p)>0$ for all $p>\bar{P}$, $f(p)<0$ for all $p<\bar{P}$.

\item Either $f(p)>0$ for all $0\le{p}\le1$, or $f(p)<0$ for all
$0\le{p}\le1$, i.e., the function $F(p)$ is either strictly
increasing or strictly decreasing.
\end{enumerate}
That is, $F(p)$ is a quasiconvex function, and there exists at most
one $p\ge0$ such that $dF(p)/dp=0$.

Note that the two conditions in \eqref{derivative} are actually
equivalent. Hence, it suffices to prove \eqref{derivative_positive}.
To this end, let us introduce the following lemma.

\begin{Lemma}\label{lemma:zeta_monotonicity}
Given $\bar{p}\ge0$ fixed, the functions $\zeta^v(p)$, which
satisfies \eqref{def:zeta_v}, and
$\zeta^e(p,\bar{p})=\zeta^e(\bar{p},p)$, which satisfies
\eqref{def:zeta_e}, are strictly decreasing for $p\ge0$, while the
functions $p\zeta^v(p)$ and $p\zeta^e(p,\bar{p})$ are strictly
increasing for $p\ge0$.
\end{Lemma}

\emph{Proof:} See Appendix \ref{sec:proof of
lemma_zeta_monotonicity}.\hfill{$\blacksquare$}

Given any $\bar{P}\ge0$ and based on Lemma
\ref{lemma:zeta_monotonicity}, we can derive a lower bound
$g(p|\bar{P})$ of $f(p)\ln2$ for all $p\ge\bar{P}$, which is given
in \eqref{long1} on the top of the next page,
\begin{figure*}
\begin{align}
f(p)\ln2&=\frac{\zeta^v(\hat{p}_{i1}^v)}{1+p\zeta^v(\hat{p}_{i1}^v)}-
\frac{\hat{p}_{i1}^v\zeta^v(p)}{1+\hat{p}_{i1}^v\zeta^v(p)}\frac{1}{\sigma^2+\sigma^2p\zeta^v(p)+p} \notag \\
&~~~~-\!\!\!\sum_{\{j|e_{ij}\in\mathcal{U}_e\}}\!\frac{\alpha_{e_{ij}}\cdot0.7\zeta^e(p,\hat{p}_{j1}^v)}{1+0.7\zeta^e(p,\hat{p}_{j1}^v)}
\frac{1+\hat{p}_{j1}^v\zeta^e(p,\hat{p}_{j1}^v)}{(1+p\zeta^e(p,\hat{p}_{j1}^v))
[\hat{p}_{j1}^v+\sigma^2(1+\hat{p}_{j1}^v\zeta^e(p,\hat{p}_{j1}^v))]+p(1+\hat{p}_{j1}^v\zeta^e(p,\hat{p}_{j1}^v))} \notag\\
&\ge\frac{\zeta^v(\hat{p}_{i1}^v)}{1+p\zeta^v(\hat{p}_{i1}^v)}-\frac{\hat{p}_{i1}^v\zeta^v(\bar{P})}
{1+\hat{p}_{i1}^v\zeta^v(\bar{P})}\frac{1}{\sigma^2+\sigma^2\bar{P}\zeta^v(\bar{P})+p} \notag \\
&~~~~-\!\!\!\sum_{\{j|e_{ij}\in\mathcal{U}_e\}}\!\frac{\alpha_{e_{ij}}\cdot0.7\zeta^e(\bar{P},\hat{p}_{j1}^v)}{1+0.7\zeta^e(\bar{P},\hat{p}_{j1}^v)}
\frac{1+\hat{p}_{j1}^v\zeta^e(p,\hat{p}_{j1}^v)}{(1+\bar{P}\zeta^e(\bar{P},\hat{p}_{j1}^v))
[\hat{p}_{j1}^v+\sigma^2(1+\hat{p}_{j1}^v\zeta^e(p,\hat{p}_{j1}^v))]+\hat{p}_{j1}^v\bar{P}\zeta^e(\bar{P},\hat{p}_{j1}^v)+p}\notag \\
&\ge\frac{\zeta^v(\hat{p}_{i1}^v)}{1+p\zeta^v(\hat{p}_{i1}^v)}-\frac{\hat{p}_{i1}^v\zeta^v(\bar{P})}{1+\hat{p}_{i1}^v\zeta^v(\bar{P})}
\frac{1}{\sigma^2+\sigma^2\bar{P}\zeta^v(\bar{P})+p} \notag \\
&~~~~-\!\!\!\sum_{\{j|e_{ij}\in\mathcal{U}_e\}}\!\frac{\alpha_{e_{ij}}\cdot0.7\zeta^e(\bar{P},\hat{p}_{j1}^v)}{1+0.7\zeta^e(\bar{P},\hat{p}_{j1}^v)}
\frac{1+\hat{p}_{j1}^v\zeta^e(\bar{P},\hat{p}_{j1}^v)}{(1+\bar{P}\zeta^e(\bar{P},\hat{p}_{j1}^v))
[\hat{p}_{j1}^v+\sigma^2(1+\hat{p}_{j1}^v\zeta^e(\bar{P},\hat{p}_{j1}^v))]+\hat{p}_{j1}^v\bar{P}
\zeta^e(\bar{P},\hat{p}_{j1}^v)+p}\notag \\
&=\frac{\zeta^v(\hat{p}_{i1}^v)}{1+p\zeta^v(\hat{p}_{i1}^v)}-g_0(p|\bar{P})-\sum_{\{j|e_{ij}\in\mathcal{U}_e\}}g_j(p|\bar{P})\triangleq{g(p|\bar{P})},~\forall{p}\ge\bar{P}\label{long1}
\end{align}
\hrulefill
\end{figure*}
To obtain the first inequality in \eqref{long1}, one can observe
that $\frac{\hat{p}_{i1}^v\zeta^v(p)}{1+\hat{p}_{i1}^v\zeta^v(p)}$
and $\frac{1}{\sigma^2+\sigma^2p\zeta^v(p)+p}$ in the second term
are increased due to the fact that $\zeta^v(p)\le\zeta^v(\bar{P})$,
$\frac{x}{1+x}$ is nondecreasing in $x$ for any $x\ge0$ and
$p\zeta^v(p)\ge\bar{P}\zeta^v(\bar{P})$; similar reasons also apply
to the summation term since
$\zeta^e(p,\hat{p}_{j1}^v)\le\zeta^e(\bar{P},\hat{p}_{j1}^v)$ and
$p\zeta^e(p,\hat{p}_{j1}^v)\ge\bar{P}\zeta^e(\bar{P},\hat{p}_{j1}^v)$
for all $p\ge\bar{P}$. The second inequality in \eqref{long1} is
obtained by the monotonicity of $\frac{x}{1+x}$ and
$\zeta^e(p,\hat{p}_{j1}^v)$. Note that the monotonic properties of
$\zeta^v(p)$, $\zeta^e(p,\hat{p}_{j1}^v)$, $p\zeta^v(p)$,
$p\zeta^e(p,\hat{p}_{j1}^v)$ and $\frac{x}{1+x}$ are strict, so the
two inequalities in \eqref{long1} hold with equality if and only if
$p=\bar{P}$; that is,
\begin{equation}\label{f_of_p_lower_bound}
\begin{cases}
f(p)\ln2>g(p|\bar{P}),& p>\bar{P},\\
f(p)\ln2=g(p|\bar{P}),& p=\bar{P}.
\end{cases}
\end{equation}

Next, we show that $g(p|\bar{P})\ge0$ for all $p\ge\bar{P}$ if
$f(\bar{P})\ge0$, which, together with \eqref{f_of_p_lower_bound}
then implies \eqref{derivative_positive}. By defining
\begin{align*}
&a_0=\sigma^2+\sigma^2\bar{P}\zeta^v(\bar{P})\ge0,\\
&b_0=\hat{p}_{i1}^v\zeta^v(\bar{P})\ge0,\\
&a_j=(1+\bar{P}\zeta^e(\bar{P},\hat{p}_{j1}^v))[\hat{p}_{j1}^v+\sigma^2(1+\hat{p}_{j1}^v\zeta^e(\bar{P},\hat{p}_{j1}^v))]\\
&~~~~~~~+\hat{p}_{j1}^v\bar{P}\zeta^e(\bar{P},\hat{p}_{j1}^v)\ge0,\\
&b_j=0.7\zeta^e(\bar{P},\hat{p}_{j1}^v)\ge0,\\
&c_j=b_j\cdot(1+\hat{p}_{j1}^v\zeta^e(\bar{P},\hat{p}_{j1}^v))\ge0,~\forall{j}\in\{k|e_{ik}\in\mathcal{U}_e\},
\end{align*}
which are independent of $p$, one can respectively express
$g_0(p|\bar{P})$ and $g_j(p|\bar{P})$ as
\begin{subequations}\label{long2}
\begin{align}
g_0(p|\bar{P})&=\frac{b_0}{(1+b_0)(a_0+p)},\label{long2_a}\\
g_j(p|\bar{P})&=\frac{\alpha_{e_{ij}}c_j}{(1+b_j)(a_j+p)},~\forall{j}\in\{k|e_{ik}\in\mathcal{U}_e\},\label{long2_b}
\end{align}
\end{subequations}
Suppose that there exists $0\le\bar{P}\le1$ satisfying
$f(\bar{P})\ge0$. Then, by \eqref{f_of_p_lower_bound}, we have
$f(\bar{P})\ln2=g(\bar{P}|\bar{P})=\frac{\zeta^v(\hat{p}_{i1}^v)}{1+\bar{P}\zeta^v(\hat{p}_{i1}^v)}-g_0(\bar{P}|\bar{P})-\sum_{\{j|e_{ij}\in\mathcal{U}_e\}}g_j(\bar{P}|\bar{P})\ge0$.
Since
$\frac{\zeta^v(\hat{p}_{i1}^v)}{1+\bar{P}\zeta^v(\hat{p}_{i1}^v)}>0$,
$g_0(\bar{P}|\bar{P})\ge0$ and $g_j(\bar{P}|\bar{P})>0$ for all
$j\in\{k|e_{ik}\in\mathcal{U}_e\}$, we can further infer from
$g(\bar{P}|\bar{P})\ge0$ that, when $p=\bar{P}$,
\begin{subequations}\label{eq:g}
\begin{align}
&\beta_0\cdot\frac{\zeta^v(\hat{p}_{i1}^v)}{1+p\zeta^v(\hat{p}_{i1}^v)}-g_0(p|\bar{P})\notag\\
&~=\frac{\big[\zeta^v(\hat{p}_{i1}^v)a_0\beta_0(1+b_0)-b_0\big]}{(1+\zeta^v(\hat{p}_{i1}^v)p)(1+b_0)(a_0+p)}\notag\\
&~~~~~~+\frac{\big[\beta_0(1+b_0)-b_0\big]\cdot\zeta^v(\hat{p}_{i1}^v)p}{(1+\zeta^v(\hat{p}_{i1}^v)p)(1+b_0)(a_0+p)}\notag\\
&~\ge0,\label{eq:g_0}\\
&\beta_j\cdot\frac{\zeta^v(\hat{p}_{i1}^v)}{1+p\zeta^v(\hat{p}_{i1}^v)}-g_j(p|\bar{P})\notag\\
&~=\frac{\big[\zeta^v(\hat{p}_{i1}^v)a_j\beta_j(1+b_j)-\alpha_{e_{ij}}c_j\big]}{(1+\zeta^v(\hat{p}_{i1}^v)p)(1+b_j)(a_j+p)}\notag\\
&~~~~~~+\frac{\big[\beta_j(1+b_j)-\alpha_{e_{ij}}c_j\big]\cdot\zeta^v(\hat{p}_{i1}^v)p}{(1+\zeta^v(\hat{p}_{i1}^v)p)(1+b_j)(a_j+p)}\notag\\
&~\ge0,~\forall{j}\in\{k|e_{ik}\in\mathcal{U}_e\},\label{eq:g_j}
\end{align}
\end{subequations}
where
\begin{align*}
&\beta_0=\frac{g_0(\bar{P}|\bar{P})}{g_0(\bar{P}|\bar{P})+\sum_{\{k|e_{ik}\in\mathcal{U}_e\}}g_j(\bar{P}|\bar{P})},\\
&\beta_j=\frac{g_j(\bar{P}|\bar{P})}{g_0(\bar{P}|\bar{P})+\sum_{\{k|e_{ik}\in\mathcal{U}_e\}}g_k(\bar{P}|\bar{P})},~{\forall}j\in\{k|e_{ik}\in\mathcal{U}_e\},
\end{align*}
and $\beta_0+\sum_{\{j|e_{ij}\in\mathcal{U}_e\}}\beta_j=1$. Next, we
aim to show that $[\beta_0(1+b_0)-b_0]\ge0$ and
$[\beta_j(1+b_j)-\alpha_{e_{ij}}c_j]\ge0$, which, when applied to
\eqref{eq:g}, imply that,
$\frac{\beta_0\zeta^v(\hat{p}_{i1}^v)}{1+p\zeta^v(\hat{p}_{i1}^v)}-g_0(p|\bar{P})\ge0$
and each
$\frac{\beta_j\zeta^v(\hat{p}_{i1}^v)}{1+p\zeta^v(\hat{p}_{i1}^v)}-g_j(p|\bar{P})\ge0$
for all $p\ge\bar{P}$. To this end, consider the following
inequalities
\begin{subequations}\label{eq:zeta_bounds}
\begin{align}
\rho{e^{\sigma^2\zeta^v}}\!(1+p\zeta^v)\!\ge&\rho(1+\sigma^2\zeta^v)(1+p\zeta^v),\label{eq:zeta_bounds_a}\\
\rho{e^{\sigma^2\zeta^e}}\!(1+p\zeta^e)(1+\hat{p}_{j1}^v\zeta^e)\!\ge&\rho(1+\sigma^2\zeta^e)(1+p\zeta^e),\label{eq:zeta_bounds_b}
\end{align}
\end{subequations}
for all $\zeta^v\ge0$, $\zeta^e\ge0$, $p\ge0$ and
$\hat{p}_{j1}\ge0$, where the inequalities are owing to $e^x\ge1+x$
for all $x$. Since $\rho=0.95<1$, for any $p\ge0$, there must exist
$\bar{\zeta}^v(p)=\bar{\zeta}^e(p)>0$ such that
\[
\rho(1+\sigma^2\bar{\zeta}^v(p))(1+p\bar{\zeta}^v(p))=\rho(1+\sigma^2\bar{\zeta}^e(p))(1+p\bar{\zeta}^e(p))=1.
\]
Besides, by \eqref{def:zeta_v}, one has
\begin{align*}
\rho{e^{\sigma^2\zeta^v(p)}}(1+p\zeta^v(p))&=1,\\
\rho{e^{\sigma^2\zeta^e(p,\hat{p}_{j1}^v)}}(1+p\zeta^e(p,\hat{p}_{j1}^v))(1+\hat{p}_{j1}^v\zeta^e(p,\hat{p}_{j1}^v))&=1.
\end{align*}
Thus, it follows from \eqref{eq:zeta_bounds_a} that
$\bar{\zeta}^v(p)\ge\zeta^v(p)$ for all $p\ge0$. Therefore, for
$0\le\bar{P}\le1$, we have
\begin{align}
\zeta^v(\hat{p}_{i1}^v)a_0&=\zeta^v(\hat{p}_{i1}^v)\sigma^2(1+\bar{P}\zeta^v(\bar{P}))\notag\\
&\le\zeta^v(0)\sigma^2(1+1\cdot\zeta^v(1))\notag\\
&\le\ln\rho^{-1}\times(1+\bar{\zeta}^v(1))\notag\\
&=\ln\rho^{-1}\!\times\!\left(\!1-\!\frac{1\!+\!\sigma^2}{2\sigma^2}\!+\!\sqrt{\frac{1}{\rho\sigma^2}\!+\!\frac{(1\!+\!\sigma^2)^2}{4\sigma^4}\!-\!\frac{1}{\sigma^2}}\right)\notag\\
&\approx0.0537<1,\label{eq:zeta_a0}
\end{align}
where the first inequality comes from Lemma
\ref{lemma:zeta_monotonicity},
$\zeta^v(0)=\frac{\ln(1/\rho)}{\sigma^2}$ is obtained from
\eqref{def:zeta_v}, and the approximate value $0.0537$ is obtained
by using $\rho=0.95$ and $\sigma^2=0.1$ in
\eqref{parameter_setting_a}. By applying \eqref{eq:zeta_a0} to
\eqref{eq:g_0}, we conclude that
\begin{equation}\label{eq:beta_0}
\beta_0(1+b_0)-b_0\ge0;
\end{equation}
otherwise \eqref{eq:g_0} does not hold true. Similarly, we have
$\bar{\zeta}^e(p)\ge\zeta^e(p,\hat{p}_{j1}^v)$, for all $p\ge0$,
$\hat{p}_{j1}^v\ge0$,
\begin{align}
&\zeta^v(\hat{p}_{i1}^v)a_j\notag\\
&~~=\zeta^v(\hat{p}_{i1}^v)\big[(1\!+\!\bar{P}\zeta^e(\bar{P},\hat{p}_{j1}^v))[\hat{p}_{j1}^v\!+\!\sigma^2(1\!+\!\hat{p}_{j1}^v\zeta^e(\bar{P},\hat{p}_{j1}^v))]\notag\\
&~~~~~~~~~~~~~~~~~~~~~~~~~~~~~~~~~~~~~~~~~~~~~~~~~~+\hat{p}_{j1}^v\bar{P}\zeta^e(\bar{P},\hat{p}_{j1}^v)\big]\notag\\
&~~\le\zeta^v(0)\big[(1\!+\!\zeta^e(1,\hat{p}_{j1}^v))[1\!+\!\sigma^2(1\!+\!{\zeta^e(\bar{P},1)})]\!+\!\zeta^e(1,\hat{p}_{j1}^v)\big]\notag\\
&~~\le\frac{1}{\sigma^2}\ln\rho^{-1}\big[(1\!+\!\bar{\zeta}^e(1))[1\!+\!\sigma^2(1\!+\!\bar{\zeta}^e(1))]\!+\!\bar{\zeta}^e(1)\big]\notag\\
&~~\approx0.6181<1,\label{eq:zeta_aj}
\end{align}
and thus \begin{equation}\label{eq:beta_j}
\beta_j(1+b_j)-\alpha_{e_{ij}}c_j\ge0,~{\forall}j\in\{k|e_{ik}\in\mathcal{U}_e\},
\end{equation}
as inferred from \eqref{eq:g_j}. By \eqref{eq:g}, \eqref{eq:beta_0}
and \eqref{eq:beta_j}, we have
\[
\frac{\beta_j\zeta^v(\hat{p}_{i1}^v)}{1\!+\!p\zeta^v(\hat{p}_{i1}^v)}-g_j(p|\bar{P})\!\ge\!0,~\text{for}~p\!\ge\!\bar{P},~j\!\in\!\{0\}\cup\{k|e_{ik}\in\mathcal{U}_e\}
\]
By summing up the above inequalities, we have $g(p|\bar{P})\ge0$ for
all $p\ge\bar{P}$. Hence, we have proved
\eqref{derivative_positive}, and the proof of Lemma
\ref{lemma:quasiconvexity} is complete.\hfill{$\blacksquare$}

\section{Proof of Lemma
\ref{lemma:zeta_monotonicity}}\label{sec:proof of
lemma_zeta_monotonicity}

Suppose that $0\!\le\!p_1\!<\!p_2$. Then, we can obtain by
\eqref{def:zeta_v} that
\begin{align*}
1&=0.95{e}^{\zeta^v(p_2)\sigma^2}(1+\zeta^v(p_2)p_2)\\
&=0.95{e}^{\zeta^v(p_1)\sigma^2}(1+\zeta^v(p_1)p_1)\\
&<0.95{e}^{\zeta^v(p_1)\sigma^2}(1+\zeta^v(p_1)p_2).
\end{align*}
Note that the above inequality is strict since $\zeta^v(p)>0$ for
any $p\ge0$. Because the function
$\rho{e}^{\zeta^v(p)\sigma^2}(1+\zeta^v(p)p)$ is strictly increasing
in $\zeta^v(p)$, the above inequality implies that
$\zeta^v(p_2)<\zeta^v(p_1)$, i.e., the function $\zeta^v(p)$ is a
strictly decreasing function for $p\ge0$. Accordingly, we can
further obtain
\begin{align*}
1&=0.95{e}^{\zeta^v(p_2)\sigma^2}(1+\zeta^v(p_2)p_2)\\
&=0.95{e}^{\zeta^v(p_1)\sigma^2}(1+\zeta^v(p_1)p_1)\\
&>0.95{e}^{\zeta^v(p_2)\sigma^2}(1+\zeta^v(p_1)p_1),
\end{align*}
which implies $\zeta^v(p_2)p_2>\zeta^v(p_1)p_1$, namely,
$p\zeta^v(p)$ is strictly increasing for $p\ge0$. The remaining
statements about $\zeta^e(p,\bar{p})$ and $p\zeta^e(p,\bar{p})$ in
Lemma \ref{lemma:zeta_monotonicity} can be proved
similarly.\hfill{$\blacksquare$}

\section*{Acknowledgement}

The authors would like to thank Prof. Che Lin of National Tsing Hua
University, Hsinchu, Taiwan, for valuable discussions in preparing
this manuscript.

\bibliographystyle{IEEEtran}
\bibliography{references}

\begin{thebibliography}{10}
\providecommand{\url}[1]{#1}
\csname url@samestyle\endcsname
\providecommand{\newblock}{\relax}
\providecommand{\bibinfo}[2]{#2}
\providecommand{\BIBentrySTDinterwordspacing}{\spaceskip=0pt\relax}
\providecommand{\BIBentryALTinterwordstretchfactor}{4}
\providecommand{\BIBentryALTinterwordspacing}{\spaceskip=\fontdimen2\font plus
\BIBentryALTinterwordstretchfactor\fontdimen3\font minus
  \fontdimen4\font\relax}
\providecommand{\BIBforeignlanguage}[2]{{%
\expandafter\ifx\csname l@#1\endcsname\relax
\typeout{** WARNING: IEEEtran.bst: No hyphenation pattern has been}%
\typeout{** loaded for the language `#1'. Using the pattern for}%
\typeout{** the default language instead.}%
\else
\language=\csname l@#1\endcsname
\fi
#2}}
\providecommand{\BIBdecl}{\relax}
\BIBdecl

\bibitem{Li2014}
W.-C. Li, T.-H. Chang, and C.-Y. Chi, ``On the complexity of {SINR} outage
  constrained max-min fairness multicell coordinated beamforming problem,'' in
  \emph{Proc. 2014 IEEE ICASSP}, Florence, Italy, May 4-9, 2014, pp.
  3508--3512.

\bibitem{Lee_2012_ComMag}
J.~Lee, Y.~Kim, H.~Lee, B.~L. Ng, D.~Mazzarese, J.~Liu, W.~Xiao, and Y.~Zhou,
  ``Coordinated multipoint transmission and reception in {LTE}-advanced
  systems,'' \emph{IEEE Commun. Mag.}, vol.~50, no.~11, pp. 44--50, Nov. 2012.

\bibitem{Gesbert10JSAC}
D.~Gesbert, S.~Hanly, H.~Huang, S.~S. Shitz, O.~Simeone, and W.~Yu,
  ``Multi-cell {MIMO} cooperative networks: A new look at interference,''
  \emph{IEEE J. Sel. Areas Commun.}, vol.~28, no.~9, pp. 1380--1408, Dec. 2010.

\bibitem{Bjornson11TSP}
E.~Bj\"{o}rnson, N.~J. Jald\'{e}n, M.~Bengtsson, and B.~Ottersten, ``Optimality
  properties, distributed strategies, and measurement-based evaluation of
  coordinated multicell {OFDMA} transmission,'' \emph{IEEE Trans. Signal
  Process.}, vol.~59, no.~12, pp. 6086--6101, Dec. 2011.

\bibitem{Annapureddy2011}
V.~S. Annapureddy and V.~V. Veeravalli, ``Sum capacity of {MIMO} interference
  channels in the low interference regime,'' \emph{IEEE Trans. Inf. Theory},
  vol.~57, no.~5, pp. 2565--2581, May 2011.

\bibitem{Liu_11}
Y.-F. Liu, Y.-H. Dai, and Z.-Q. Luo, ``Coordinated beamforming for {MISO}
  interference channel: {Complexity} analysis and efficient algorithms,''
  \emph{IEEE Trans. Signal Process.}, vol.~59, no.~3, pp. 1142--1157, Mar.
  2011.

\bibitem{Liu_13}
------, ``Max-min fairness linear transceiver design for a multi-user {MIMO}
  interference channel,'' \emph{IEEE Trans. Signal Process.}, vol.~61, pp.
  2413--2423, May 2013.

\bibitem{Jorswieck08}
E.~A. Jorswieck, E.~G. Larsson, and D.~Danev, ``Complete characterization of
  the {Pareto} boundary for the {MISO} interference channel,'' \emph{IEEE
  Trans. Signal Process.}, vol.~56, no.~10, pp. 5292--5296, Oct. 2008.

\bibitem{Mochaourab_11}
R.~Mochaourab and E.~A. Jorswieck, ``Optimal beamforming in interference
  networks with perfect local channel information,'' \emph{IEEE Trans. Signal
  Process.}, vol.~59, no.~3, pp. 1128--1141, Mar. 2011.

\bibitem{Zakhour_09}
R.~Zakhour and D.~Gesbert, ``Coordination on the {MISO} interference channel
  using the virtual {SINR} framework,'' in \emph{Proc. Int. ITG Workshop on
  Smart Antennas}, Berlin, Germany, Feb. 16-18, 2009.

\bibitem{Bjornson_2012_Pareto}
E.~Bjornson, M.~Bengtsson, and B.~Ottersten, ``Pareto characterization of the
  multicell {MIMO} performance region with simple receivers,'' \emph{IEEE
  Trans. Signal Process.}, vol.~60, no.~8, pp. 4464--4469, 2012.

\bibitem{MAPEL}
L.~P. Qian, Y.~Zhang, and J.~Huang, ``{MAPEL}: {Achieving} global optimality
  for a non-convex wireless power control problem,'' \emph{IEEE Trans. Wireless
  Commun.}, vol.~8, no.~3, pp. 1553--1563, Mar. 2009.

\bibitem{Jorswieck2010_POA}
E.~A. Jorswieck and E.~G. Larsson, ``Monotonic optimization framework for the
  two-user {MISO} interference channel,'' \emph{IEEE Trans. Commun.}, vol.~58,
  no.~7, pp. 2159--2168, July 2010.

\bibitem{Utschick2012_POA}
W.~Utschick and J.~Brehmer, ``Monotonic optimization framework for coordinated
  beamforming in multicell networks,'' \emph{IEEE Trans. Signal Process.},
  vol.~60, no.~4, pp. 1899--1909, Apr. 2012.

\bibitem{Zhang_2012_POA}
L.~Liu, R.~Zhang, and K.-C. Chua, ``Achieving global optimality for weighted
  sum-rate maximization in the {$K$}-user {Gaussian} interference channel with
  multiple antennas,'' \emph{IEEE Trans. Wireless Commun.}, vol.~11, no.~5, pp.
  1933--1945, May 2012.

\bibitem{Schmidt_09}
D.~A. Schmidt, C.~Shi, R.~A. Berry, M.~L. Honig, and W.~Utschick, ``Distributed
  resource allocation schemes: {Pricing} algorithms for power control and
  beamformer design in interference networks,'' \emph{IEEE Signal Process.
  Mag.}, vol.~26, no.~5, pp. 53--63, Sep. 2009.

\bibitem{Zhang_Cui_2010}
R.~Zhang and S.~Cui, ``Cooperative interference management with {MISO}
  beamforming,'' \emph{IEEE Trans. Signal Process.}, vol.~58, pp. 5450--5458,
  Oct. 2010.

\bibitem{Kim_2011}
S.-J. Kim and G.~Giannakis, ``Optimal resource allocation for {MIMO} ad hoc
  cognitive radio networks,'' \emph{IEEE Trans. Inf. Theory}, vol.~57, pp.
  3117--3131, May 2011.

\bibitem{Shi2011_IteMMSE}
Q.~Shi, M.~Razaviyayn, Z.-Q. Luo, and C.~He, ``An iteratively weighted {MMSE}
  approach to distributed sum-utility maximization for a {MIMO} interfering
  broadcast channel,'' \emph{IEEE Trans. Signal Process.}, vol.~59, no.~9, pp.
  4331--4340, Sep. 2011.

\bibitem{Nguyen_11}
D.~H.~N. Nguyen and T.~Le-Ngoc, ``Multiuser downlink beamforming in multicell
  wireless systems: {A} game theoretical approach,'' \emph{IEEE Trans. Signal
  Process.}, vol.~59, pp. 3326--3338, July 2011.

\bibitem{Hong2012}
M.-Y. Hong and Z.-Q. Luo, ``Signal processing and optimal resource allocation
  for the interference channel,'' \emph{Academic Press Library in Signal
  Process., arXiv:1206.5144v1}, 2013.

\bibitem{Weeraddana_2013}
P.~Weeraddana, M.~Codreanu, M.~Latva-aho, and A.~Ephremides, ``Multicell {MISO}
  downlink weighted sum-rate maximization: {A} distributed approach,''
  \emph{IEEE Trans. Signal Process.}, vol.~61, no.~3, pp. 556--570, 2013.

\bibitem{Lindblom_11}
J.~Lindblom, E.~Karipidis, and E.~G. Larsson, ``Outage rate regions for the
  {MISO} interference channel: {Definitions} and interpretations,''
  \emph{http://arxiv.org/abs/1106.5615v1}.

\bibitem{Park2012}
J.~Park, Y.~Sung, D.~Kim, and H.~V. Poor, ``Outage probability and outage-based
  robust beamforming for {MIMO} interference channels with imperfect channel
  state information,'' \emph{IEEE Trans. Wireless Commun.}, vol.~11, pp.
  3561--3573, June 2012.

\bibitem{Li_13}
W.-C. Li, T.-H. Chang, C.~Lin, and C.-Y. Chi, ``Coordinated beamforming for
  multiuser {MISO} interference channel under rate outage constraints,''
  \emph{IEEE Trans. Signal Process.}, vol.~61, pp. 1087--1103, Mar. 2013.

\bibitem{Kandukuri02}
S.~Kandukuri and S.~Boyd, ``Optimal power control in interference-limited
  fading wireless channels with outage-probability specifications,'' \emph{IEEE
  Trans. Wireless Commun.}, vol.~1, pp. 46--55, Jan. 2002.

\bibitem{Ghosh_10}
S.~Ghosh, B.~D. Rao, and J.~R. Zeidler, ``Outage-efficient strategies for
  multiuser {MIMO} networks with channel distribution information,'' \emph{IEEE
  Trans. Signal Process.}, vol.~58, pp. 6312--6324, Dec. 2010.

\bibitem{Tan_2011}
C.~W. Tan, ``Optimal power control in {Rayleigh}-fading heterogeneous
  networks,'' in \emph{Proc. IEEE INFOCOM}, Shanghai, April 10-15, 2011, pp.
  2552--2560.

\bibitem{Huang_2012_GLOBECOM}
Y.~Huang, C.~W. Tan, and B.~Rao, ``Outage balancing in multiuser {MISO}
  networks: {Network} duality and algorithms,'' in \emph{Proc. IEEE GLOBECOM},
  Anaheim, CA, Dec. 3-7, 2012, pp. 3918--3923.

\bibitem{Karp2010_21NPComplete}
R.~M. Karp, ``Reducibility among combinatorial problems,'' in \emph{50 Years of
  Integer Programming 1958-2008}, M.~J\"{u}nger, T.~M. Liebling, D.~Naddef,
  G.~L. Nemhauser, W.~R. Pulleyblank, G.~Reinelt, G.~Rinaldi, and L.~A. Wolsey,
  Eds.\hskip 1em plus 0.5em minus 0.4em\relax Springer Berlin Heidelberg, 2010,
  ch.~8, pp. 219--241.

\bibitem{Li_14_alg}
\BIBentryALTinterwordspacing
W.-C. Li, T.-H. Chang, and C.-Y. Chi, ``Multicell coordinated beamforming with
  rate outage constraint--{Part II}: {Efficient} approximation algorithms,''
  \emph{submitted to IEEE Trans. Signal Process.}, 2014. [Online]. Available:
  \url{http://arxiv.org/abs/1405.2984}
\BIBentrySTDinterwordspacing

\bibitem{Luo_Zhang2008}
Z.-Q. Luo and S.~Zhang, ``Dynamic spectrum management: Complexity and
  duality,'' \emph{IEEE J. Sel. Topics Signal Process.}, vol.~2, pp. 57--73,
  Feb. 2008.

\bibitem{Li2011}
W.-C. Li, T.-H. Chang, C.~Lin, and C.-Y. Chi, ``A convex approximation approach
  to weighted sum rate maximization of multiuser {MISO} interference channel
  under outage constraints,'' in \emph{Proc. 2011 IEEE ICASSP}, Progue, Czech,
  May 22-27, 2011, pp. 3368--3371.

\bibitem{BK:BoydV04}
S.~Boyd and L.~Vandenberghe, \emph{Convex {O}ptimization}.\hskip 1em plus 0.5em
  minus 0.4em\relax Cambridge, UK: Cambridge University Press, 2004.

\bibitem{Krantz_Parks02}
S.~G. Krantz and H.~R. Parks, \emph{The Implicit Function Theorem: History,
  Theory, and Applications}.\hskip 1em plus 0.5em minus 0.4em\relax Boston, MA:
  Birkh\"{a}user, 2002.

\end{thebibliography}

\end{document}